\documentclass{article}
\usepackage[letterpaper, margin=1in]{geometry}

\usepackage[hyphens]{url}
\usepackage[hidelinks]{hyperref}

\usepackage{comment}
\usepackage{graphicx}
\usepackage{listings}
\usepackage{amsmath}

\usepackage{booktabs} 

\usepackage[linesnumbered,ruled]{algorithm2e}
\SetKwRepeat{Do}{do}{while}
\SetCommentSty{mycommfont}

\usepackage[frozencache,cachedir=.]{minted}

\usepackage{authblk}

\newcommand{\code}[1]{\textit{#1}}  

\newcounter{example}[section]
\newenvironment{example}[1][]{\refstepcounter{example}\par\medskip
   \noindent \textbf{Example~\theexample. #1} \rmfamily}{\medskip}

\begin{document}
\title{Immutable Log Storage as a Service on Private and Public Blockchains}

\author{William Pourmajidi$^1$, Lei Zhang$^2$, John Steinbacher$^3$, Tony Erwin$^4$,  and Andriy Miranskyy$^5$ \\
$^{1,2,5}$ Department of Computer Science, Ryerson University, Toronto, Canada \\
$^{3}$ IBM Canada Lab, Toronto, Canada \\
$^{4}$ IBM Watson and Cloud Platform, Austin, USA \\
william.pourmajidi@ryerson.ca, leizhang@ryerson.ca, jstein@ca.ibm.com, aerwin@us.ibm.com, avm@ryerson.ca}

\date{}

\maketitle
\begin{abstract}
Service Level Agreements (SLA) are employed to ensure the performance of Cloud solutions. When a component fails, the importance of logs increases significantly. All departments may turn to logs to determine the cause of the issue and find the party at fault. The party at fault may be motivated to tamper with the logs to hide their role. We argue that the critical nature of Cloud logs calls for immutability and verification mechanism without the presence of a single trusted party.

This paper proposes such a mechanism by describing a blockchain-based log storage system, called Logchain, which can be integrated with existing private and public blockchain solutions. Logchain uses the immutability feature of blockchain to provide a tamper-resistance platform for log storage. Additionally, we propose a hierarchical structure to address blockchains' scalability issues. To validate the mechanism, we integrate Logchain into Ethereum and IBM Blockchain. We show that the solution is scalable and perform the analysis of the cost of ownership to help a reader select an implementation that would address their needs.

The Logchain's scalability improvement on a blockchain is achieved without any alteration of blockchains' fundamental architecture. As shown in this work, it can function on private and public blockchains and, therefore, can be a suitable alternative for organizations that need a secure, immutable log storage platform.

\end{abstract}

\section{Introduction}\label{sec:Introduction}

In the majority of Cloud offerings, there are two parties involved. A Cloud Service Provider (CSP) owns a pool of computing resources and offers them at predefined prices to a Cloud Service Consumer (CSC) via the Internet.

The CSP uses continued monitoring to ensure that the current Quality of Service (QoS) provided to the CSC matches with the one in the signed Service Level Agreement (SLA). When a technical issue arises, the CSC will also become interested in reviewing and assessing the logs, because logs hold the truth about the delivered QoS. 

While the full control over monitoring systems allows CSPs to monitor Cloud services efficiently, it gives them a controversial power over evidential resources essential to CSCs. That is, logs are generated and stored on a platform, which is built, managed, and owned by a CSP. Hence, CSPs have full permission on all collected logs. Such situations cause many trust-related issues.

Logs are evidential documents~\cite{accorsi2009log}. They contain all the details and QoS metrics related to the operation of software, network components, servers, and Cloud platforms. As a key element in computer forensic investigations, logs are presentable in the court of law~\cite{selamat2008mapping, reilly2010Cloud} only if they satisfy requirements, such as authenticity and reliability.

Log tampering include adding, removing, and manipulating a log partially or entirely. We provide three log tampering examples in Appendix~\ref{sec:examples}. Moreover, log tampering may affect CSCs financially and technically. If a CSP tampers with the logs related to resource usage and overcharges the customer, or if a CSP hides the breach of one or more criteria of an SLA, the CSC is in immediate need of finding a method or a tool to verify the integrity of data provided by the CSP.

Given the existence of many motivations to tamper with logs, the negative consequences of log tampering for the CSCs, and the inadequacy of existing monitoring solutions, we conclude that an immutable log system, which is capable of storing the logs and verifying their integrity, can be genuinely beneficial for the CSCs and can be used to establish trust among Cloud participants.

To address the trust issue among Cloud participants, our \textbf{objective} is to create an immutable log system called Logchain (LC). We choose blockchain as our immutable  storage option. The proposed solution collects logs from various platforms and stores them in blocks of blockchains. To make the LC more accessible, we implement an API module for LC that allows participants to submit logs to the LC and verify the integrity of the submitted logs to the LC. As the API allows the LC to be used as a service, from here onward, we use the term Logchain as a service (LCaaS) to refer to the API-enabled LC.

Traditionally, blockchains use two types of blocks. The genesis blocks mark the start of a blockchain, and the data blocks store data in the blockchain. Additional details about these blocks are provided in Section~\ref{subsec:Common Key Components of Blockchains}. We introduce a conceptual model in which we extend the definition of the data blocks to allow us to construct additional types of blocks while keeping the internal architecture of the data blocks intact. We design a special type of data block named Terminal Block (TB). The terminal block terminates a blockchain, and the terminated blockchain can no longer accept data blocks. The definition and technical details of additionally constructed block types are provided in Section~\ref{subsec:Enhancements on Blockchain Structure}. In our definition of special types of blocks, we keep the internal architecture of data blocks intact to ensure blockchains can work with LCaaS in the future.

The LCaaS API receives the logs (or their digest) and sends them to LCaaS~\code{Blockify} module to convert them to the blocks and store the blocks in the blockchain. Blockchains are known to have scalability issues~\cite{chauhan2018blockchain}. Hence, to address the scalability issues of blockchains, we build a hierarchical ledger that uses inter-related blockchains to form a scalable log storage platform. In LCaaS, blocks are constantly added to a lower-level blockchain until a time- or size-bound threshold is triggered. Once the trigger is sent to the~\code{Blockify} module, it needs to terminate the current lower-level blockchain, summarize its important data into a TB, and submit the content of the TB to a data block of a higher-level blockchain. TB contains the most important information needed for verification of data stored in all the blocks of a lower-level blockchain that it has terminated. Therefore, verifying the integrity of a single TB proves that none of the blocks of a lower-level blockchain have been tampered with, saving time and resources on the verification process. The details of LCaaS design are captured in Section~\ref{sec:Design of Lcaas}.

The main \textbf{contribution} of this paper is to extend our work\footnote{In~\cite{pourmajidi2018logchain,pourmajidi2019immutable}, we mainly focused on design of LCaaS and its interaction with Ethereum blockchain.} on the applicability of LCaaS as a hierarchical ledger~\cite{pourmajidi2018logchain, pourmajidi2019immutable} from a proprietary proof-of-concept blockchain to public and private commercial blockchains. We selected Ethereum~\cite{ethereum}  as a public blockchain vendor and  IBM~\cite{IBMBlockchain} as a private blockchain vendor. We chose them as they are popular and representative blockchains vendors.

We compare the results of LCaaS integration to these two blockchain platforms. We also assess and compare the cost of ownership for the two implementations.

We focus on the primary integration point between LCaaS and blockchain vendors and assess integrated solutions' performance by measuring the impact of all relevant factors. The incoming transactions per second ($tps$) defines the incoming load for the LCaaS. The higher the $tps$, the more work there is for LCaaS and its hashing methods. Nevertheless, the $tps$ does not directly impact the number of outgoing transactions to blockchain vendors. The number of outgoing transactions is defined by the length of lower-level blockchains, also known as circled blockchains. The longer circled blockchains result in a lower frequency of submission of blocks to blockchain vendors. While this lower submission makes the integration more feasible (as there are fewer transactions to be paid for), it creates a longer window of time for logs to be stored locally before they can be sent to blockchain vendor, hence a higher risk. 

While there are solutions that recommend the use of blockchain for storage of critical data (e.g.,~\cite{shafagh2017towards, chen2019blockchain, xu2018blockchain}), to the best of our knowledge, none of them have tried to address the scalability issues of blockchains for large-scale log storage. As a result, most existing solutions have settled for partial storage of logs, such as audit logs~\cite{kelsey1999minimizing, waters2004building}. While this solves a portion of the problem, we argue that the entire log storage system has to be immutable.

We design the proposed solution as a service so that multiple customers can use it concurrently. Moreover, given that CSCs are already using many ``X-as-a-Service'' solutions, they are accustomed to such a service offering model. 

The clients of LCaaS can use its API to interact with the solution and submit and verify logs. The source code of LCaaS can be accessed via~\cite{LCaaS}.

CSPs can use LCaaS to promote visibility, trust, and accountability. CSPs can choose to submit their operational logs to LCaaS and offer a verification mechanism to their CSCs. Similarly, the CSCs can confirm the authenticity of the generated logs for their consumed services using the verification API. The CSPs can decide what components or portions of their infrastructure need to send logs to the LCaaS. Ideally, all the Cloud platform components, from the core network to the user-facing services, should submit their operational logs to the LCaaS. The more components of the Cloud platform that submit logs to the LCaaS, the more trustable the relationship between the CSPs and the CSCs will become. However, as submission of data to blockchain incurs a cost, a justifiable feasibility assessment is suggested. Additionally, suppose CSPs are concerned about the confidentiality and secrets that submission of the actual logs to the LCaaS may reveal. In that case, they can choose to submit the digest of the logs. While submission of the digest of logs offers peace of mind in terms of any potential breach of privacy or sensitive information, LCaaS can not offer the original version of the logs if they are tampered with.

The rest of this paper is structured as follows. In Section~\ref{sec:Literature Review}, we provide related works and a brief literature review on both Cloud computing and blockchain.  
In Section~\ref{sec:Design of Lcaas}, we introduce the methodologies that we use to build LCaaS, our prototype, and the approaches for enhancing blockchain and its capability, followed by the implementation of the hierarchical ledger. 
In Section~\ref{sec:Validation Case Study}, we present implementation and integration details of LCaaS to blockchain vendors, namely, Ethereum and IBM Blockchain. In Section~\ref{sec:Results}, we analyze the results of the integration of the LCaaS model with Ethereum and IBM Blockchain. In Section~\ref{sec:discussions}, we discuss practical aspects of implementing LCaaS. In Section~\ref{sec:Threats to Validity}, we summarize potential threats to the validity of the proposed solution. Finally, in Section~\ref{sec:Conclusion}, we conclude the paper by providing a summary and a direction towards future work.

\section{Literature Review}\label{sec:Literature Review}

In Section~\ref{subsec:Cloud Logs and Related Challenges}, we review the importance of Cloud monitoring and its related challenges.
In Section~\ref{subsec:blockchain}, we provide an overview of blockchain and its characteristics, concluding with a review of blockchain capacity limitations and their solutions. We also refer the reader to Appendix~\ref{sec:Background} for background details on blockchain and its taxonomies.

\subsection{Cloud Logs and Related Challenges}\label{subsec:Cloud Logs and Related Challenges}

The legal system relies on a range of forensic investigation and identification. In the digital era, digital evidence such as operational logs, transaction logs, and usage logs replace physical evidence. This new type of evidence requires new forensic investigation methods~\cite{farid2009image}. In an attempt to regulate and standardize digital forensics practices, the Digital Forensic Research Workshop (DFRWS)~\cite{DFRWS} has developed a forensic framework. The framework identifies preservation as a crucial step and indicates that it must be a guarded principle across ``forensic'' categories. 

Another major challenge related to the authenticity and reliability of digital evidence is that digital data are much more easily tampered with compared to physical evidence. 
To effectively use logs as digital evidence, many have recommended using a Log Management System (LMS) ~\cite{Marty:2011:CAL:1982185.1982226,ray2013secure}. The majority of LMSs promise a set of desirable features, such as tamper-resistance, verifiability, confidentiality, and privacy~\cite{ray2013secure}. 
To secure the collected logs, specifically audit logs, Waters et al.~\cite{waters2004building} build a platform that uses hash encryption to protect the audit logs from unauthorized parties by encrypting the content of them. 
However, LMSs are managed by IT personnel; thus, requiring the presence of a trusted third party (TTP), limiting LMSs applicability.

In recent years, many verification-as-a-service platforms offer integrity control for the data that are uploaded by the user, but they do need a TTP. For example, arXiv~\cite{arXiv} provides a repository that ensures document's integrity.

The problem of trusting a third-party can be alleviated by a self-contained solution that does not rely on a TTP integrity verification service. We argue that blockchain is among the most promising solutions that can be used to replace the requirement for a TTP. It was initially designed to support a cryptocurrency known as Bitcoin that does not require a TTP (such as banks or other financial institutes) for verification or maintenance of financial transactions. A correctly implemented distributed blockchain is an adequate alternative to address the TTP issue~\cite{Underwood2016Blockchain, mainelli2015sharing}.

\subsection{Blockchain} \label{subsec:blockchain}

A significant component of blockchain that allows transactions to be immutable is the use of the Proof of Work (PoW) verification schema. PoW involves running iterations for finding a particular value that, when it is hashed in conjunction with other elements of a block, the calculated hash begins with a certain number of zero bits. The number of zeros is proportional to the time required to complete a PoW. The higher the number of zeros, the longer it will take to complete the PoW. Once the computational effort is dedicated and the hash value is found, all items along with the found value, known as \code{nonce}, are kept in a block. The content of a block cannot be changed unless the whole PoW process is repeated. Chaining blocks together using hash binding or hash chaining~\cite{kelsey1999minimizing,haber1990time,schneier1998cryptographic} significantly increases the amount of computational effort that is needed for changing the content of an earlier block. In a hash binding relationship, the current hash of the previous block is used as the previous hash of the current block. This chain makes any attempt to change the blockchain computationally unfeasible as one needs to re-process PoW for all the blocks in order to tamper with any of the earlier blocks~\cite{matzutt2018thwarting}.
 
Current blockchain implementations of the distributed ledgers already have notary proof-of-existence services~\cite{Underwood2016Blockchain}. For example, Poex.io~\cite{ProofofExistence}, launched in 2013, verifies the existence of a computer file at a specific time, by storing a timestamp and the SHA-256~\cite{gallagher2008secure} of the respective file in a block that eventually will be added to a blockchain. 
Proof-of-existence solutions cannot be used as scalable LMSs, as they consider files individually, with no search function to locate the appropriate file or block.
Moreover, Cloud solutions consist of thousands of components, each of which generates a large volume of logs~\cite{pourmajidi2017challenges}. The current solutions are not designed to handle the scale that is required to store Cloud-generated logs. Furthermore, the current public blockchains support limited concurrent transactions~\cite{Underwood2016Blockchain}.

Although blockchain technology has great potential and can be used in many disciplines, it is dealing with a number of challenges. The scalability remains the most critical challenge~\cite{zheng2016blockchain}. Blockchain heavily relies on consensus algorithms, like PoW, and such algorithms are computationally expensive. To overcome the scalability issues, a novel cryptocurrency scheme is suggested by~\cite{brucemini} where old transactions are removed from the blockchain, and a database holds the values of removed transactions. Although this solution reduces the size of the blockchain, it introduces the same trust issue that traditional databases are suffering from. 
 
Eyal et al.~\cite{eyal2016bitcoin} suggest redesigning the current structure of a blockchain. In the redesigned model, known as Bitcoin-NG (next generation), conventional blocks are decoupled into two parts: the key block and microblocks. The key block is used for leader election, and the leader is responsible for microblock generation until a new leader appears.

\section{Design of LCaaS}\label{sec:Design of Lcaas}

In this section, we introduce the methodologies that we use to build LCaaS and explore the implementation of the hierarchical ledger. Here, we provide a general overview of LCaaS, its details, and the additional components that are added to the blockchain to enhance LCaaS' scalability. LCaaS API, its signatures and methods are given in Appendix~\ref{subsec:LCaaS API}.

Current blockchain consensus protocols require every node of the network to process every blockchain block, hence a major scalability limitation. PoW is among the most common consensus algorithms at the time of writing this work~\cite{bach2018comparative}. Thus, to ensure that our proposed solution is useful for the majority of blockchains, we use PoW as the main consensus algorithm for the LCaaS. However, our modular design allows one to easily replace PoW with any other consensus algorithms such as Proof-of-Stake, Proof-of-Importance, or newer ones based on the Byzantine Fault Tolerance (BFT), which aim to solve the problem of reaching consensus when nodes can generate arbitrary data~\cite{bouraga2021taxonomy}. 

As blockchain consists of various components, different solutions have been offered for scalability issues of each specific component. For instance, On-chain, Off-chain, and Inter-chain solutions are used to improve the overall performance of blockchains~\cite{kim2018survey}. LCaaS is mainly concerned with the performance issues caused by the size of the blockchain (number of blocks) and uses an approach known as Child-chain~\cite{pawar2021study}, where a parent-child structure is promoted and continuous transactions are stored in a child-chain (similar to our lower-level blockchain), and the results are stored in a parent-chain (similar to our high-level blockchain). 

The performance of a blockchain platform is mainly impacted by the number of its blocks, the size of these blocks, and the consensus algorithm it uses. While choosing an appropriate consensus algorithm has a major impact on the blockchain performance ~\cite{bamakan2020survey}, our focus is on the performance issues caused by a large number of blocks in a never-ending blockchain.

We overcome this size issue by segmenting a portion of a blockchain and locking it down in a block of a higher level blockchain, i.e., we create a two-level hierarchy of blockchains. Validating the integrity of a high-level block confirms the integrity of all the blocks of the lower-level blockchain. This segmentation leads to a more efficient validation process. The hierarchical structure of LCaaS is graphically shown in Appendix~\ref{subsec:Hierarchical Structure of LCaaS}.

LCaaS resides on top of a basic blockchain and converts it to a hierarchical ledger. Our primary goal is to bring scalability to blockchain for the situations in which the number of data items stored in a blockchain is large (e.g., operational logs of a Cloud platform). 

As the name implies, the LCaaS offers the hierarchical ledger as a service. Cloud participants can create an account and receive a unique API key for all corresponding API calls. Clients also need to configure two main settings on their instance before they can use it. The first key configuration is the \code{difficulty\_target}, which is defined as the number of required zeros at the beginning of an acceptable hash. The LCaaS will continue to generate new hashes and new \code{nonces} until a hash is generated that matches the difficulty target. The second key configuration is defining a limit for the number of blocks in a circled blockchain. This constraint acts as a size-limit and controls how many blocks are accepted in each circled blockchain. Once the limit is reached, the LCaaS takes the blockchain and pushes it to the hierarchical ledger. Let us now look at the key components of the LCaaS.

\subsection{Enhancements on Blockchain Structure}\label{subsec:Enhancements on Blockchain Structure}

While common key components of blockchains are necessary to implement a blockchain, LCaaS requires additional components. We have introduced absolute genesis block, relative genesis block, terminal block, circled blockchains, super block, and super blockchains. These advancements allow the LCaaS to provide the hierarchical structure that improves the scalability of blockchains.

\textit{Absolute Genesis Block (AGB):} Similar to the Markle root in a Markle tree~\cite{tasca2017taxonomy}, the absolute genesis block is placed as the first block of the first circled blockchain. An AGB is the first block that is created in the LCaaS and has the same characteristics as GB, with \code{index} and \code{previous\_hash} set to zero and the \code{data} element set to null. 

\textit{Relative Genesis Block (RGB):} Relative genesis block is placed at the beginning of every subsequent circled blockchain after the first circled blockchain. The \code{previous\_hash} of an RGB is set to the \code{current\_hash} of the terminal block of the previous circled blockchain.

\textit{Terminal Blocks (TB):} Terminal Blocks are added at the end of a blockchain to ``close'' it and produce a circled blockchain. The terminal block's \code{index} and \code{current\_hash} are calculated similarly to any other block. The part that differentiates a terminal block from a genesis block or a data block is its \code{data} element. Data blocks of blockchains are designed to be data-type agnostic. The majority of Cloud monitoring tools allow data to be submitted to external sources for further analysis. At the time of writing this work, JSON format is the de facto data exchange standard for such data exchange. Hence, we assume that the CSP can submit data to our API using the JSON format. As a result, the terminal block's \code{data} element stores a JSON object that contains details about the terminated circled blockchain. Practitioners who adopt LCaaS may choose other machine-readable formats, such as XML, YAML, or a user-defined one. The details and the processes to create a TB are as follows. The \code{aggr\_hash} is created by collecting and hashing \code{current\_hash} values of all blocks in that circled blockchain, from the AGB or RGB to the block before the terminal block. The \code{data} element also store four additional values, namely \code{timestamp\_from}, \code{timestamp\_to}, \code{block\_index\_from}, and \code{block\_index\_to}.

\textit{Circled Blockchains (CB):} Circled blockchains are blockchains that are capped. In other words, there is a limit on the number of blocks that they can include before a terminal block ``caps'' the blocks. Once a circled blockchain is terminated, it can not accept any new block.

\textit{Super Blocks (SB):} Super blocks exhibit the features of regular data blocks and have \code{nonce}, \code{index}, \code{timestamp}, \code{data}, \code{previous\_hash}, and \code{current\_hash}. The only difference between a super block and data block is that super block's \code{data} element stores all of the field of a terminal block of a circled blockchain. In order to accept terminal block elements, the \code{data} element consists of a JSON object. The elements of this JSON object are as follows: \code{index}, \code{timestamp}, \code{data}, \code{current\_hash}, \code{previous\_hash}, and \code{nonce}.

\textit{Super Blockchain (SBC):} Super blockchain is a blockchain where each of its blocks is a super block. The super blocks are ``chained'' together by hash binding. In other words, super blocks that are linked together will result in a super blockchain. An~$i$-th super block in a super blockchain relies on \code{current\_hash} of its previous super block. If data in an earlier super block $m$ is tampered with, the link among all the subsequent super blocks, from $m+1$ to the most recent super block, denoted by super block $i$, will be broken. Then one has to recompute \code{current\_hash} and \code{nonce} values of each super block from super block $m$ to super block $i$.

Figure~\ref{fig::TerminalBlockdetails} shows the relationship between a TB and all other blocks in a CB. All elements of a SB are identical to the ones of a data block. Thus, it can be implemented by any other blockchain framework.

Considering that a SB's data element includes all the elements of a TB, changing any block in a CB, not only breaks the CB but also breaks the SBC. 
The relationship between a TB and the data element of a SB is shown in Appendix~\ref{subsec:Hierarchical Structure of LCaaS}.

The data element of the SB provides a hash tree structure and enhances the immutability the CBs, while decreasing the computational resources required to verify blocks in a circled blockchain. 

The above novel enhancements allow the LCaaS to provide the hierarchical structure that is needed to overcome scalability
limitations of the blockchains.

\section{Validation Case Study}\label{sec:Validation Case Study}

Here, we review the validation of the LCaaS.  Section~\ref{subsec:Integration Platforms - Rationale} provides rationale for choosing Ethereum and IBM Blockchain. Sections~\ref{subsec:Case study setup: Ethereum} and~\ref{subsec:Case study setup: IBM Blockchain} cover details of integration with the Ethereum and IBM Blockchains, respectively.
Section~\ref{subsec:Control Factors} discusses controlling factors of Ethereum and IBM Blockchain integration. Finally, Section~\ref{subsec:Workload Drivers} depicts the workload drivers.

\subsection{Integration Platforms - Rationale}\label{subsec:Integration Platforms - Rationale}

\subsubsection{Rationale for choosing Ethereum}\label{subsubsec:Rationale for choosing Ethereum}

While the focus of some public blockchains (such as Bitcoin and Litecoin) is on financial transactions, other public blockchains such as Ethereum try to provide different use cases for blockchains. Coherently, Ethereum provides the developer with an end-to-end system for building various distributed applications~\cite{wood2014ethereum}. 

The smart contracts are autonomous pieces of code~\cite{szabo1997idea}. They are deployed over the blockchain and upon being called, can interact with the user data or data stored in the blockchain. Ethereum has developed Solidity~\cite{solidity}, a high-level language for smart contracts, and Remix~\cite{remix}, an Integrated Development Environment (IDE) for Solidity.

Given the popularity of Ethereum, we have selected Ethereum as the public blockchain platform for integration with the LCaaS.

\subsubsection{Rationale for choosing IBM Blockchain}\label{subsubsec:Rationale for choosing IBM Blockchain}

IBM Blockchain provides developers with an end-to-end platform for designing, building, and implementing various applications based on the underlying blockchain technology~\cite{Blockcha89:online}. 

Another major feature of IBM Blockchain is its extensive support for the smart contracts. Within the IBM Blockchain, smart contracts are known as Chaincode~\cite{Glossary38:online}. To simplify the development process, IBM Blockchain is equipped with an extension~\cite{IBMBlock61:online} for Visual Studio code(VSC)~\cite{VisualSt19:online} that can be used for building and deploying smart contract on the IBM Blockchain platform. The extension uses direct access to IBM Blockchain (of course, after successful authentication) and can directly edit, run, and deploy smart contracts on the IBM Blockchain. Smart contracts for IBM Blockchain can be developed in Java, JavaScript, and Go~\cite{Developi20:online}.  

Given its popularity and its wide range of use cases~\cite{Blockcha89:online}, we have selected IBM Blockchain as the private blockchain platform for integration with the LCaaS.

\subsection{Case study setup: Ethereum}\label{subsec:Case study setup: Ethereum}

\begin{figure}[t]
\centering
\includegraphics[width=\columnwidth]{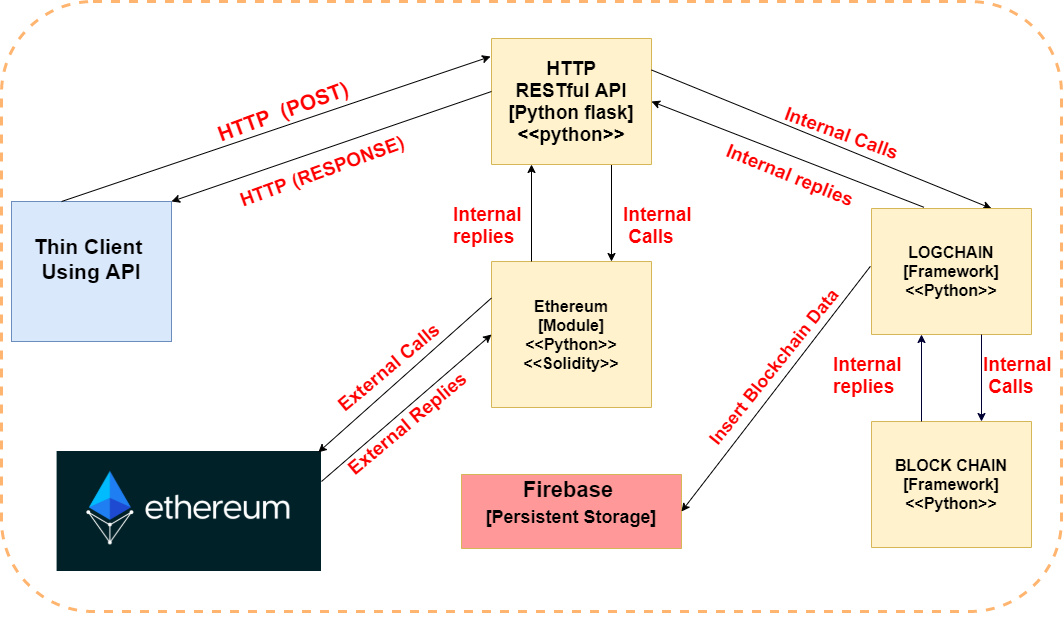}
\caption{LCaaS and Ethereum integration}
\label{fig:LCaaS and Ethereum Integration}
\end{figure}

Figure~\ref{fig:LCaaS and Ethereum Integration} depicts the relationship between LCaaS and Ethereum. The \code{API} module receives API calls from clients and passes them to the \code{Logchain} Module. The \code{Logchain} module employs the \code{Blockchain} module to convert the received data (digest or raw logs) to blocks and pushes a copy of the blocks to Firebase and another copy to the \code{API} module. The \code{API} module pushes the received data to the Ethereum and informs the client of the successful submission of data to the blockchain. We use Google Firebase real-time database~\cite{firebase} to store all types of blocks as additional permanent storage. Integration with the Firebase can be disabled without affecting the normal operation of the LCaaS. 

LCaaS is built on top of a private blockchain. In order to replace it with a public blockchain (e.g., Ethereum), integration points have to be designed. We propose a composite structure, in which receiving logs and converting them to blocks happens at the LCaaS side and storing the hashes and digitally signing them happens over the Ethereum blockchain. Using blockchain terminology, data collection and blockification of logs happen off-chain, and the block storage on the Ethereum blockchain is handled by Ethereum smart contract and will be on-chain.

Within the Ethereum blockchain, economics is controlled by an execution fee called gas. The gas is paid by Ether---the Ethereum intrinsic currency~\cite{wood2014ethereum}. Gas measures, in computation resource terms, the effort that is needed to process the transaction. A smart contract consists of one or more operations, and each operation has an associated gas cost, which is defined by the Ethereum protocol~\cite{AccountT87:online}. For instance, a SHA-3 operation costs 30 units of gas. The higher the gas price, the more appealing the transaction would become for the miners. Hence, if a transaction needs to be executed faster, the higher gas price will motivate a miner to consider the transaction and mine it in the upcoming block. The current implementation of LCaaS does not use a lot of computational resources for each block that is pushed to the blockchain; thus, the main bottleneck is at the blockchain provider side. As indicated above, one can increase the performance of the blockchain by increasing the gas price for the desired transaction.

In the light of the above economics and the fact that each transaction incurs a cost, we limited the submissions to the Ethereum blockchain to super blocks. Super blocks include complete elements of a terminal block of a circled blockchain and can be used to verify the integrity of all the blocks in that circled blockchain. Based on the preimage resistance property of hash functions (mentioned in Section~\ref{subsec:MiningAndHashBinding}), it would be computationally infeasible to construct an entire circled blockchain such that its hash matches the \code{current\_hash} of a super block. However, if one desires, minor changes to the current implementation of the \code{Ethereum} module can be made to allow the LCaaS to push data blocks to the Ethereum blockchain as well.

Since the main Ethereum blockchain is used for production, Ethereum team has allocated a test network for development purposes~\cite{ethereumtestnet}. In this work, we have used the Ethereum test network (as opposed to the real one) for the integration of the LCaaS and Ethereum. The cost of transactions on the Ethereum test network is paid with a special type of Ether, known as test Ether which does not have any real monetary value. To obtain test Ether, we use MetaMask Ether Faucet~\cite{faucet}.

Smart contracts on the Ethereum blockchain allow users to interact with the blockchain. The storage of SBs in the blockchain requires a smart contract. It is important to mention that a SB's data element contains the terminal block of a CB. Hence, SB is the most efficient candidate to be stored in a public blockchain, as one can easily verify the integrity of a SB and conclude the integrity of all the blocks in the CB that the TB has terminated. 

We use Solidity~\cite{solidity} to develop the smart contract and name it~\code{Superblock.sol} (available at~\cite{LCaaS}). Once the smart contract is developed, it has to be published on the Ethereum blockchain. We use the Remix to publish the smart contract. Publishing a smart contract on the blockchain is considered a transaction and is a chargeable service. Thus, we use the test Ether that we have stored in MetaMask vault to pay the transaction fee. 

The published smart contract stores the SBs on the Ethereum blockchain and, upon successful submission, returns a receipt back to the \code{Ethereum} module. The receipt includes details of the transaction, such as the sender address, the content, the transaction hash, and the block number.

All interactions with the Ethereum blockchain can be traced using Etherscan~\cite{etherscan}, a web dashboard connected to Ethereum blockchain. Known as Ethereum block explorer, Etherscan allows anyone to look up transactions' details by using the sender or recipient address, transaction hash, or block number. 

Using the Ethereum \code{blockNumber} field or the sender address, one can verify the transaction. If the transaction is found and the submitted SB matches with the one at hand, the integrity of the TB in the data element of the SB is confirmed; thus, confirming the integrity of all blocks in the CB that the TB has terminated. An example of a successful transaction of LCaaS on Etherscan can be seen in~\cite{RopstenT14:online}.

\subsection{Case study setup: IBM Blockchain}\label{subsec:Case study setup: IBM Blockchain}

\begin{figure}[t]
\centering
\includegraphics[width=\columnwidth]{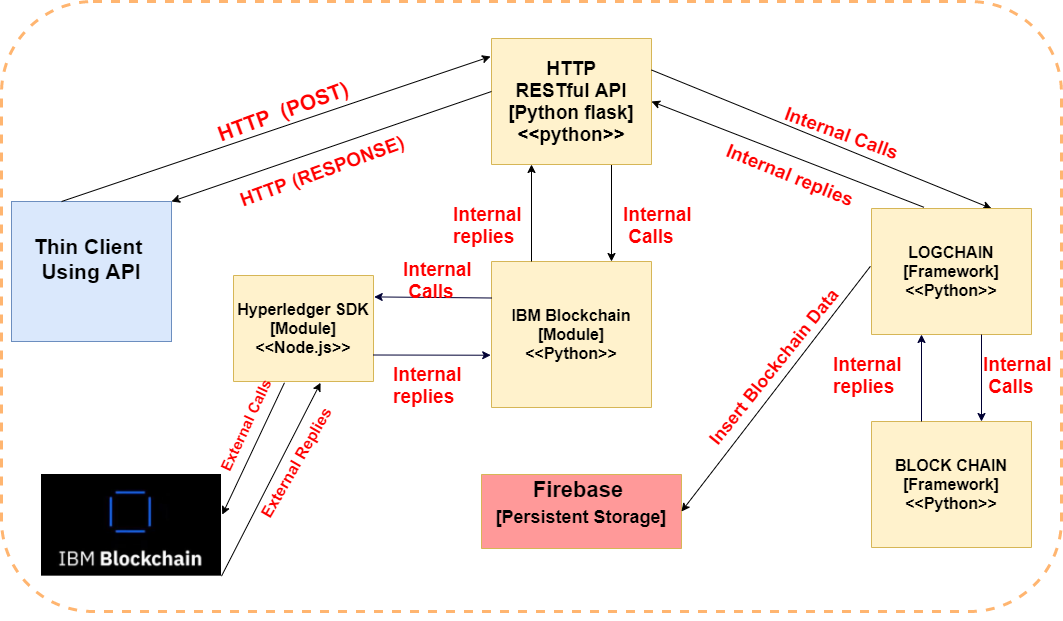}
\caption{LCaaS and IBM Blockchain integration}
\label{fig:LCaaS and IBM Blockchain Integration}
\end{figure}

The relationship between LCaaS and IBM Blockchain is depicted in Figure~\ref{fig:LCaaS and IBM Blockchain Integration}. 
The process of receiving data, processing it, and converting it to blocks remain the same as the processes explained in Section~\ref{subsec:Case study setup: Ethereum} except that data are pushed to IBM Blockchain instead of Ethereum.

As for the integration between LCaaS and IBM Blockchain,  receiving logs and converting them to blocks happens at the LCaaS side and storing the hashes and digitally signing them happens over the IBM Blockchain. 

Unlike Ethereum, there is no concept of gas price on IBM Blockchain. However, the economics are controlled, on hourly bases, and based on virtual processor core (VPC) allocation. This simplified model is based on the amount of CPU (or VPC) that the IBM Blockchain Platform nodes are allocated on an hourly basis. To further clarify the concept of VPC, is important to mention that a VPC is a unit of measurement that is used to determine the licensing cost of IBM products and is based on the number of virtual cores (vCPUs) that are available to the product. A vCPU is a virtual core that is assigned to a virtual machine or a physical processor core. Within the IBM Blockchain platform, the platform cost estimation for 1 VPC = 1 CPU = 1 vCPU = 1 Core~\cite{Pricingf4:online}.

There are some benefits of using this pricing model for IBM Blockchain. To begin with, it resembles Cloud pricing mode, namely hourly, pay-as-you-go model. Additionally, it brings clarity to estimations and as there is no minimum-investment requirement, developers and early adopters can try out the platform 

Given that this is a paid service where each transaction incurs a cost, we limited the submissions to the IBM Blockchain to super blocks as they include complete elements of a terminal block of a circled blockchain and can be used to verify the integrity of all the blocks in that circled blockchain. Figure~\ref{fig:LCaaS and IBM Blockchain Integration} depicts the relationship between LCaaS and IBM Blockchain.

The IBM Blockchain is a paid-service, and can be deployed on the IBM Cloud~\cite{IBMCloud87:online} as a service. We create an account on the IBM Cloud and follow the guidelines listed on the IBM Blockchain developer support website~\cite{Getstart38:online} to deploy and configure an instance of the IBM Blockchain. The latest version of VSC was downloaded and the IBM Blockchain add-on for the VSC was installed. As reviewed in Section~\ref{subsubsec:Rationale for choosing Ethereum}, smart contracts allow external actors to interact with a blockchain. In the case of IBM Blockchain, smart contracts, known as chaincodes, are used for the same purpose. 

The IBM Blockchain uses Hyperledger fabric~\cite{Hyperled77:online} as its blockchain framework and work based on the open source tools hosted by the Linux Foundation~\cite{Hyperled6:online}. Thus, we created a local development/deployment environment similar to the IBM Blockchain by instantiating a local version of the Hyperledger fabric blockchain. Once the smart contract code was up and running, we ported it to the IBM Blockchain. 

In this work, to develop smart contracts, our first attempt was to implement them using Python; however, the Hyperledger Python SDK~\cite{GitHubhy46:online} was not ready at the time of this experiment; hence, we chose JavaScript (running on the Node.js backend). The next step was to package the smart contract using the IBM Blockchain Platform Extension for VSC and export it so it can be imported to the IBM Blockchain network. 

The VSC IBM Blockchain extension allows direct interaction with the IBM Blockchain as long as a private blockchain with credentials (permissioned) is set up, and its connection details are configured on the VSC side. To instantiate an instance of The IBM Blockchain platform, one has to create a network~\cite{Buildane6:online} onto an IBM Kubernetes~\cite{Kubernet36:online}. After deploying the network, we created the required Membership Service Provider (MSP) and a Certificate Authority(CA) for the private blockchain and associated them with a channel. Each transaction on a network is executed on a channel that is a private sub-network of the main blockchain and only allowed participants (known as organizations) can communicate and conduct transactions. Once the channel is created, the developed chaincode is imported and associated with it. The deployed chaincode constructs the business logic that receives data from LCaaS and submits it to the IBM Blockchain. At this stage, the IBM Blockchain is configured, and there is a channel with authorized participants who can submit data as blocks to the private instance of IBM Blockchain instantiated on the IBM Cloud. The only remaining part for the end-to-end integration is a module that connects LCaaS to the instantiated instance of IBM Blockchain as the private blockchain. 

Many distributed blockchains, such as Ethereum and Bitcoin, are not permissioned, which means that any node can participate in the consensus process, wherein transactions are ordered and bundled into blocks. Because of this fact, these systems rely on probabilistic consensus algorithms which eventually guarantee ledger consistency to a high degree of probability, but which are still vulnerable to divergent ledgers (also known as a ledger “fork”), where different participants in the network have a different view of the accepted order of transactions.
As for the Hyperledger Fabric, the open source platform behind the IBM Blockchain, things work differently. Hyperledger Fabric features a kind of a node called an orderer (it’s also known as an “ordering node”) that does this transaction ordering, which along with other nodes forms an ordering service. Because Fabric’s design relies on deterministic consensus algorithms, any block a peer validates as generated by the ordering service is guaranteed to be final and correct. Ledgers cannot fork the way they do in many other distributed blockchains.

\subsection{Test bench and control factors}\label{subsec:Control Factors}

\subsubsection{Test bench}
To test the performance of LCaaS and its integration with blockchain vendors, we designed a load test which we run on our test computer with Intel i7-7500U CPU and 16 GB of RAM. The main goal of the load test is to evaluate the impact of the configurable factors: incoming transactions per second and length of circled blockchains.

For Ethereum, we use publicly available Ethereum test network (as was discussed in Section~\ref{subsec:Case study setup: Ethereum}).

IBM Blockchain can be deployed in two different environments. The first, and preferred method is to deploy it on the IBM Cloud while the second option is to install and configure it on-premises~\cite{Blockcha47:online}. We chose the Cloud deployment as it was the recommended deployment model. The default configuration of computing resources of the IBM Blockchain~\cite{Pricingf47:online} seem adequate for our test scenarios, but one can increase the computing resources for a larger storage and/or a higher performance. 

To be able to compare the results of submission of super blocks to Ethereum and IBM Blockchain, we kept the rest of the configurations alike. Table~\ref{tbl:control_factors} provides a summary of controlling factors and workload drivers for Ethereum and IBM blockchain respectively. Below, we will provide the details of each of the factors.

\begin{table}[tb]
    \centering
    \caption{Configurable factors. Note that $g$ is specific to Ethereum integration; $1 \textrm{Ether} = 10^{9} \textrm{gwei}$.}\label{tbl:control_factors}
    \begin{tabular}{ll}
    \toprule
         Factor & Values \\
    \midrule
         Transactions per second ($tps$) &  $[0.1, 1, 10, 100]$  \\
         Length of circled blockchain ($n$) &  $[1, 10, 100]$ \\
         Gas price ($g$) measured in gwei      & $[6,9,20]$ \\
         Number of sample log files     & $200$ or $1000$ \\
         Size of sample log files measured in bytes      & $64$ \\
         \code{difficulty\_target} measured in bytes & $000$ prefix \\
   \bottomrule
   \end{tabular}
\end{table}

\subsubsection{Common configurable factors}
Incoming transactions per second ($tps$): The number of $tps$ depends on the velocity of log generation at the CSP side and the way that CSP decides to push their logs into LCaaS. The transaction in this case may represent a log file. It is important to mention that the change in $tps$ does not directly lead to a change of frequency of submissions to the Ethereum blockchain. This is because LCaaS will store the received logs in blocks of a circled blockchain and then will push the generated super block to the Ethereum. By varying the value of $tps$, we mimic workloads with different intensities, from the least intense at $tps=0.1$ to the most intense at $tps=100$. 

Length of circled blockchains ($n$): Circled blockchains store a genesis block, one or many data blocks, and one terminal block. In other words, the number of data blocks in circled blockchains is configurable. A larger number of data blocks in a circled blockchain will result in a lower frequency of submissions to the Ethereum. Since each submission has a transaction fee, configuring a larger value for the number of data blocks in a circled blockchain seems more feasible. While this is true, it brings an additional risk. The risk is caused by the fact that LCaaS stores generated blocks internally and waits until the circled blockchain is terminated before sending the respective super block to the Ethereum blockchain. Hence, transactions in a longer circled blockchain are stored for a longer period, and this gives a longer window to an adversary to tamper with the logs (we will further discuss this in Section~\ref{sec:security}). Therefore, by varying the value of $n$, we mimic LCaaS handling from the most sensitive data (at $n=1$) to the least sensitive one (at $n=100$). 
We submit the digest of a log file ($64$-byte long) using \code{submit\_digest} function. Thus, the length of the file is always constant. We did not measure the performance of \code{submit\_raw} in this series of experiments, but we know that most of the time in this function is spent in computing the digest, which is proportional to the length of the file. On our test computer, it takes 200 ms to compute SHA-256 digest for 1 MB file, 1.5 seconds --- for 10 MB file, and 15 seconds --- for 100 MB file. The time to transfer the raw file from the user to LCaaS will also be proportionate to the length of the file. However, if the internal network is fast, then the transfer time will be small and can be ignored. Note that the time needed to process a super block is independent of the length of a raw file, as we are dealing with digests of the files at that stage.

We set the difficulty target for internal computations to $000$. To ensure that there are enough submissions to Ethereum for each of the 36 experiments, for the setups with $n \in [1, 10]$, we submit $200$ digests representing $200$ log files to the LCaaS; and for the setups with $n=100$, we submit $1000$ digests representing $1000$ log files to the LCaaS.

\subsubsection{Ethereum-specific control factors}
\label{subsubsec:Ethereum integration - Control factors} 

In addition to $tps$ and $n$, Ethereum has gas price ($g$) as another controlling factor. 

Gas price ($g$): The gas price is the amount paid per unit of gas and is defined by the initiator of the transaction. The higher the gas price, the more appealing the transaction would become for the miners. Hence, if a transaction needs to be executed faster, the higher gas price will motivate a miner to consider the transaction and mine it in the upcoming block. The ETH Gas Station~\cite{ETHGasSt50:online} keeps track of all submitted transactions and their processing times and suggests a value of $g$ for different processing speeds. For our experiments, we tried setting $g \in [6, 9, 20]$ gwei. These three values of $g$ on September 10, 2018 corresponded to processing times of less than 30, 5, and 2 minutes, respectively (based on~\cite{ETHGasSt50:online}).

\subsubsection{Test scenarios}

Based on the configurable items and suggested values for gas price\footnote{Suggested gas prices as per~\cite{ETHGasSt50:online}, gathered on Sep. 10, 2018 were as follows. For SafeLow tier executed in ($<$ 30 minutes)~--- 6 gwei; for Standard tier ($<$ 5 minutes)~--- 9 gwei, and for Fast tier ($<$ 2 minutes)~--- 20 gwei.}, for Ethereum, we designed 36 scenarios consist of combination of each one of the following possible values:
$tps = [0.1, 1, 10, 100]$, $n= [1, 10, 100]$, and $g = [6, 9, 20]$.

Unlike Ethereum, IBM Blockchain does not have any currency or gas price, hence, we deal with a subset of the controlling factors, namely, $tps$ and $n$, leading to 12 distinct setups.

\subsection{Workload Drivers}\label{subsec:Workload Drivers}

Since the LCaaS receives its incoming data through API, we use Postman~\cite{PostmanA97:online} to generate incoming transactions (i.e., log files) to the LCaaS. Using the \code{Runner} function of Postman, one can set the number of iterations and delay for each submitted API calls. For example, the number of iterations set to $200$ and the delay of $1000$ milliseconds will send $200$ transactions to LCaaS at the rate of $1$ $tps$.

\section{Results}\label{sec:Results}

In Section~\ref{subsec:LCaaS performance test analysis}, we discuss the performance of the the LCaaS itself, comparing the processing time for each block type. In Sections~\ref{subsec:Ethereum integration performance test analysis} and \ref{subsec:IBM Blockchain integration performance test analysis}, we discuss integration performance with Ethereum and IBM Blockchain, respectively. 

\subsection{LCaaS performance test analysis}\label{subsec:LCaaS performance test analysis}

LCaaS component, integrated with Ethereum and IBM Blockchain, is kept identical in both integration scenarios (so that we can fairly assess the performance of each integration). That is, the number of transactions, the length of circled blockchains, and the size of submitted logs to LCaaS are identical in both integration scenarios. Therefore, the only difference between the two setups, from the LCaaS point of view, is the blockchain vendor that is chosen as the endpoint for submissions of super blocks. 

\begin{figure*}[ht]
    \centering
    \includegraphics[width=0.85\textwidth]{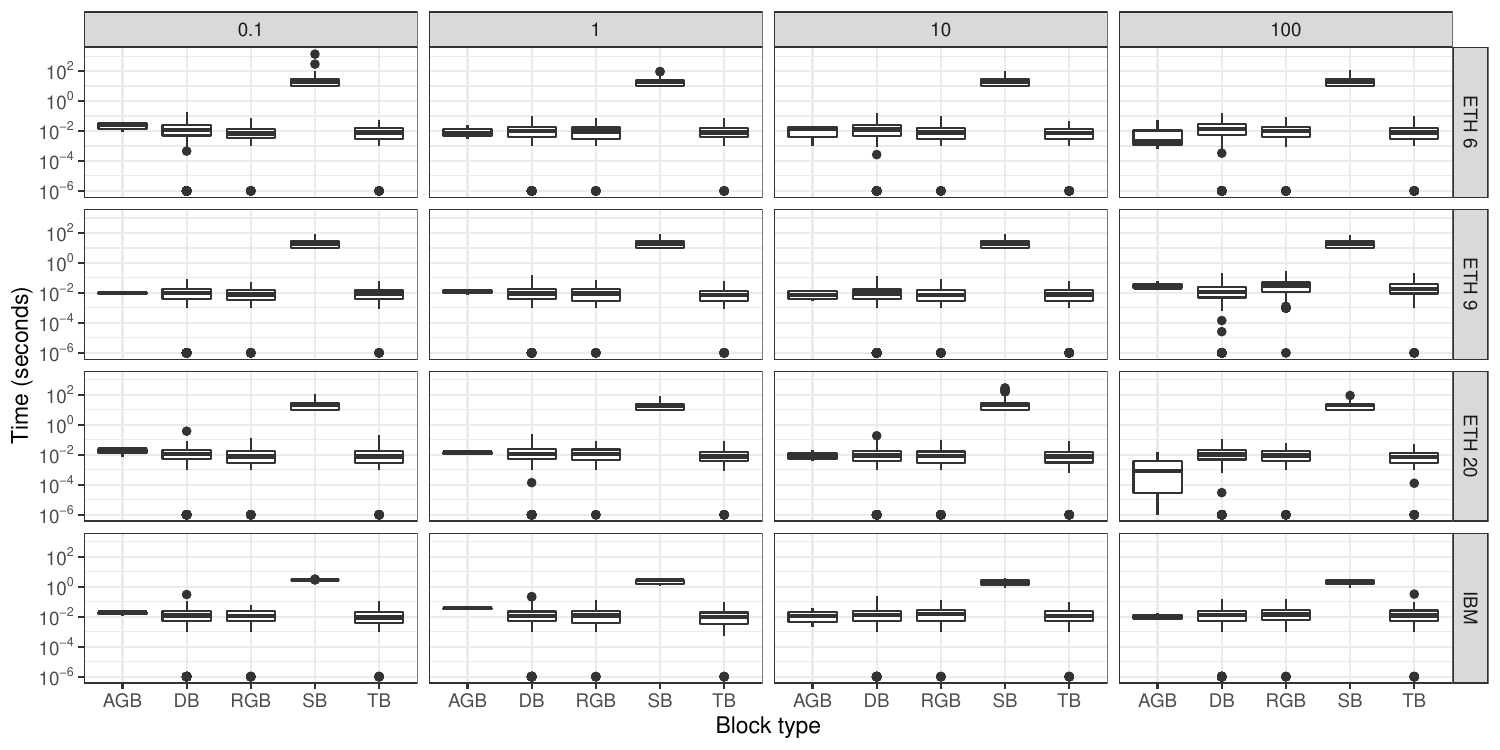}
    \caption{Processing time of different types of blocks for different experiments. Columns portray different values of $tps$, namely, 0.1, 1, 10, and 100. Rows show different gas prices in the case of Ethereum LCaaS implementation (denoted ETH), namely, 6, 9, and 20, and no gas price in the case of IBM LCaaS implementation (denoted IBM).}
    \label{fig:perf_plots}
\end{figure*}

Internally, we track the time needed to create a block of each block type (discussed in Section~\ref{sec:Design of Lcaas}). The results of the timing for each block type are listed in Figure~\ref{fig:perf_plots} and Table~\ref{tab:summary_stats}. As expected, by construction, the creation of internal blocks on LCaaS (namely, AGB, DB, RGB, and TB) is much faster than of the SB, which has to be submitted to blockchain vendors. Internal blocks creation time ranges\footnote{Technically, the $10^{-6}$ should be interpreted as $\leq 10^{-6}$, as $10^{-6}$ is the smallest duration that we could measure.} between $10^{-6}$ and 0.38 seconds, while creation of the SB --- from 10.17 seconds to 23 minutes for Ethereum and from 0.78 to 3.63 seconds for IBM Blockchain.

The time to create an internal block depends mainly on the \code{difficulty\_target} and the processing speed of a computer on which the test is performed. This is why the processing times for all types of internal blocks are similar. Moreover, eyeballing of Figure~\ref{fig:perf_plots} suggests that the processing time of the AGB, DB, RGB, and TB blocks remains similar, independent of the values of $tps$ and $n$. This suggests that our test bed is capable to absorb both low- and high-intensity workloads without reaching the saturation.

To create a super block, one needs to parse and fetch all of the field of a terminal block of a circled blockchain and send it out to an external vendor, hence the additional processing time. Let us look at the SB submission time in details.

\begin{table}[ht]

\caption{Summary statistics for processing time of different types of blocks. The columns of `1st~qu.' and `3rd~qu.' denote first and thirds quantiles, respectively. SB (ETH) and SB (IBM) denote super blocks submitted to Ethereum and IBM Blockchain, respectively.}
\label{tab:summary_stats}
\centering
\begin{tabular}{lrrrrrr}
\toprule
Block & min & 1st qu. & median & mean & 3rd qu. & max\\
\midrule
AGB & 1E-6 & 0.01 & 0.01 & 0.02 & 0.02 & 0.06\\
DB & 1E-6 & 0.00 & 0.01 & 0.02 & 0.02 & 0.38\\
RGB & 1E-6 & 0.00 & 0.01 & 0.02 & 0.02 & 0.31\\
TB & 1E-6 & 0.00 & 0.01 & 0.01 & 0.02 & 0.32\\
SB (ETH) & 10.17 & 10.33 & 20.34 & 23.10 & 30.44 & 1353.37\\
SB (IBM) & 0.78 & 1.62 & 2.72 & 2.28 & 2.79 & 3.63\\
\bottomrule
\end{tabular}
\end{table}

\subsection{Ethereum integration performance test analysis}\label{subsec:Ethereum integration performance test analysis}

We conduct 36 experiments (one for every permutation of the values of the factors listed in Table~\ref{tbl:control_factors}). 

To see whether the performance will be affected by $n$, $g$ and $tps$, we performed Pearson and Spearman correlation analysis as well as linear regression analysis on the raw data (i.e., per SB timing), the mean, the median, and the 95th percentile timing\footnote{Aggregate statistics are chosen to reduce the amount of noise in the data.} of SBs for each experiment. We found that none of the factors or the composite factors have any statistically significant relation to the response times, based on the low ($<0.15$) values of correlations and high ($>0.1$) $p$-values of linear models. This implies that the time needed to process SB block is dependent mainly on external factors, e.g., saturation of the Ethereum network and availability of the miners.

\begin{figure}[ht]
    \centering
    \includegraphics[width=0.75\columnwidth]{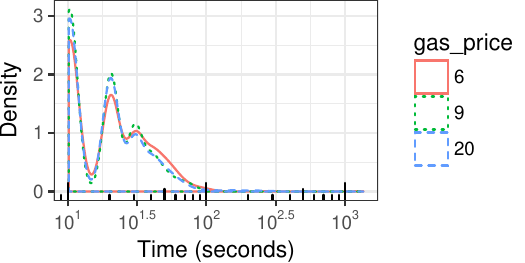}
    \caption{Density estimate of the processing time for SB block on Ethereum test network.}
    \label{fig:Ethereum_density_estimtes}
\end{figure}

We show a distribution of processing times for SB block in Figure~\ref{fig:Ethereum_density_estimtes}. Eyeballing of the distributions suggests that the lower the gas price is, the more SB blocks have higher processing time ($> 32$ seconds), even though the difference is not dramatic. Based on Kolmogorov-Smirnov test, the distribution of $g = 20$ case differs significantly ($p$-value $< 0.001$) from the cases when $g=9$ or $g=6$. However, the difference between $g=6$ and $g=9$ cases is less pronounced: $p$-value $\approx 0.08$. We were anticipating a stronger difference between all three cases; probably our usage of the test network rather than a production one lead to this anemic difference. 

The Ethereum production network strives to keep the average processing time of a transaction at $\approx$ 15 seconds~\cite{etherscan}. As shown in Table~\ref{tab:summary_stats}, we observe an average of 23 seconds and the median of 20 seconds, making it close to the target (even on the test network). Essentially, our findings show that the network has enough capacity to ``absorb'' the changes in our workload even in the intense cases, such as $tps = 100$ and $n=1$. However, in rare cases, the processing time is high: out of 3089 processed SBs, 5 (0.16\%) had been processed in between 3 and 5 minutes, and 1 (0.03\%) in 23 minutes. As we can see, these cases are rare, but they do exist and we have to be aware of such events. From practical perspective, if LCaaS workload deals with sensitive data, one may setup an alerting mechanism that will notify stakeholders of such events.

Our tests show the overall feasibility of the approach: the smart contract required to submit an SB is low enough ($\approx$335K units of gas, as shown in a sample SB submission~\cite{RopstenT14:online}) to fit into an Ethereum block (i.e., it can be executed on the production network). We also know that the production Ethereum network is scalable (currently handling $\approx$1.2 million transactions per day~\cite{Ethereum42:online}). Thus, adding hundreds or thousands of SB submissions per day will not saturate the network. 

The processing time is strongly driven by gas prices on the production network, as shown empirically by~\cite{ETHGasSt50:online}. We will further discuss gas prices and their effect on the cost of ownership in Section~\ref{sec:disc_eth}--onward.

\subsection{IBM Blockchain integration performance test analysis
}\label{subsec:IBM Blockchain integration performance test analysis}

As the concept of gas price does not apply to the IBM Blockchain integration, our controlling factors and their subsequent scenarios are limited to 12 experiments (one for every permutation of the values of the factors listed in Table~\ref{tbl:control_factors}).

For each experiment, we have introduced submission timers for each block that is submitted to the IBM Blockchain. To measure the most relevant timer, we have enabled this timer on the Hyperledger Node.js component of the IBM Blockchain integration (see Figure~\ref{fig:LCaaS and IBM Blockchain Integration}). 

As shown in Figure~\ref{fig:perf_plots} and Table~\ref{tab:summary_stats}, the submission time of an SB ranges between 0.78 and 3.63 seconds. Kolmogorov-Smirnov test shows that $tps$ and $n$ affect the duration of the submission time. However, from practical perspective, their impact is minor as the submission time, even in the worst-case scenario is less than 4 seconds.

For the IBM Blockchain, the main factor controlling the SB processing time is the performance of underlying hardware running the IBM Blockchain. Furthermore, as IBM Blockchain is implemented over Kubernetes containers, additional clustering configuration can be used to scale up the Blockchain platform performance, if needed. The results published by the IBM Blockchain team show that the platform can handle 128 peers and 325 channels using LevelDB ledger and achieve around 13K $tps$~\cite{DoesHype22:online}.

\section{Discussion}\label{sec:discussions}

In the previous section, we have analyzed the results of integration between LCaaS and two blockchain vendors, namely, Ethereum and IBM Blockchain. We choose these two vendors on purpose as each one represents a different and vital type of blockchain. The private blockchains are designed for businesses that want to employ blockchain technologies but without a publicly accessible ledger. While this can be an applicable scenario for many businesses, many others, such as online asset tracking systems, need publicly available ledgers, hence the public blockchains. 

Which solution is the best? As usual, it depends on various aspects of a particular use-case. Below, we will discuss these aspects. Specifically, implementation of the LCaaS part of our solution in production is discussed in Section~\ref{sec:lcaas_production}, security --- in Section~\ref{sec:security}, timing --- in Section~\ref{sec:timing}, and the cost of ownership --- in Section~\ref{sec:costs}.

\subsection{Platform for CBs}\label{sec:lcaas_production}
We built a prototype implementing core elements (covering CBs) of the LCaaS~\cite{LCaaS} ourselves. However, for the production implementations, we recommend that practitioners build their solution on top of the enterprise-grade blockchain services, such as the IBM Blockchain that was used in this work, or similar solutions, such as AWS Blockchain~\cite{Blockcha84:online} and Azure Blockchain~\cite{Blockcha8:online}. For the private implementation, one can use one of the Hyperledger frameworks, such as Hyperledger Fabric~\cite{Fabric-Hyperledger}, and build a hierarchical ledger on top of it. Furthermore, public blockchain services, such as Ethereum, can be used but may not be financially feasible for a large number of logs. Essentially, one will need to create a single blockchain for Super blockchain and additional ones for each of the CBs. Note that a personal instance of the Ethereum network can be deployed on the Cloud. For example, Microsoft Azure Cloud provides a template that can create the infrastructure needed to deploy components needed for the creation of the network~\cite{AzureBlo74:online}. 

\subsection{Security}\label{sec:security}
\subsubsection{Log submission intensity}
How should we set a policy of submitting the logs to the LCaaS? Two control variables at our disposal are the time interval between submissions to the LCaaS and the maximum length (or the number of records) that a log file has to reach before it gets submitted to LCaaS.
Intuitively, if we submit a log file to LCaaS every week, then it will give a perpetrator sufficient time to alter this log. Submitting logs every second (or every time a new log record comes in) will mitigate this risk, but will probably be economically infeasible and may also lead to scalability challenges. 

What is the ``goldilocks zone'' then? The answer will depend on the nature of the logs. 
Log files with sensitive information, such as security and audit logs (e.g., recording event of a user logging into a system) may have to be submitted to LCaaS individually, i.e., a log file would contain a single log record. Typically, the intensity of arrival of such events is low and will not overwhelm computational resources and budget. Moreover, these are the types of records that an analyst may want to preserve as-is, without hashing them (assuming this does not violate confidentiality).
Log files with less sensitive information, e.g., the operational logs containing performance metrics, can be submitted to the system every six hours (or sooner in case the log gets full and is truncated and archived by an operating system, based on policies set by IT personnel). In this case, four transactions per day will be submitted by a single logger, making archiving scalable and budget-friendly. Note that these are the types of records that are good candidates for hashing rather than storing in the raw format.

\subsubsection{Public vs. private blockchain}

The security of blockchain implementations is mainly dependent upon the security of underlying software and hardware as well as the protocols and settings required for the blockchain to function~\cite{gao2018survey}.

A public blockchain is more decentralized, with a large number of participating nodes. In contrast, a private blockchain is more centralized, and is designed to be used by one or more groups of users with a common goal and inherently, a fairly smaller number of participating nodes. The number of participating nodes, the consensus protocol in place, and the type of implementation play a significant role in security aspects of blockchain-based solutions~\cite{agbo2019comparison}

\subsection{Timing}\label{sec:timing}
As we can see from Figure~\ref{fig:perf_plots} and Table~\ref{tab:summary_stats}, the processing of a transaction on average takes two seconds on IBM Blockchain, while on the Ethereum network, it takes at least ten seconds.

For IBM Blockchain, we are using the production network and thus can readily use these statistics. However, in the case of Ethereum, we are using the test network. For a proper comparison, we need to assess the time required to sign a transaction on the Ethereum blockchain.

The timing on the production network will depend on multiple factors, such as the number of miners and the size of their computing resources, the intensity and the size of the transactions submitted to the network. All of these factors are outside of our control.

The Ethereum production network strives to keep the average processing time of a transaction at $\approx$15 seconds~\cite{etherscan}. We can treat it the lowest time limit to process and sign a block. In practice, the lowest threshold is a bit higher, as a user needs to spend time to send the transaction to the pool of the transactions and then distribute the transaction between the miners. Thus, signing an SB is faster on the IBM Blockchain than on the Ethereum Network. Let us now look at the cost of ownership of these two solutions.

\subsection{Cost of ownership}\label{sec:costs}

The costs associated with the creation and maintenance of the CBs will be the same, independent of the solution that we build to sign the SBs. Let us examine the costs of signing the SBs below.

\subsubsection{Ethereum Blockchain}\label{sec:disc_eth}

In the case of Ethereum, SB submissions do not incur fixed cost, as we do not need any specialized infrastructure for Ethereum. However, the cost will depend on the number of transactions submitted (as we have to pay for every transaction) and the price of gas.

\begin{figure}[bt]
    \centering
    \includegraphics[width=0.8\columnwidth]{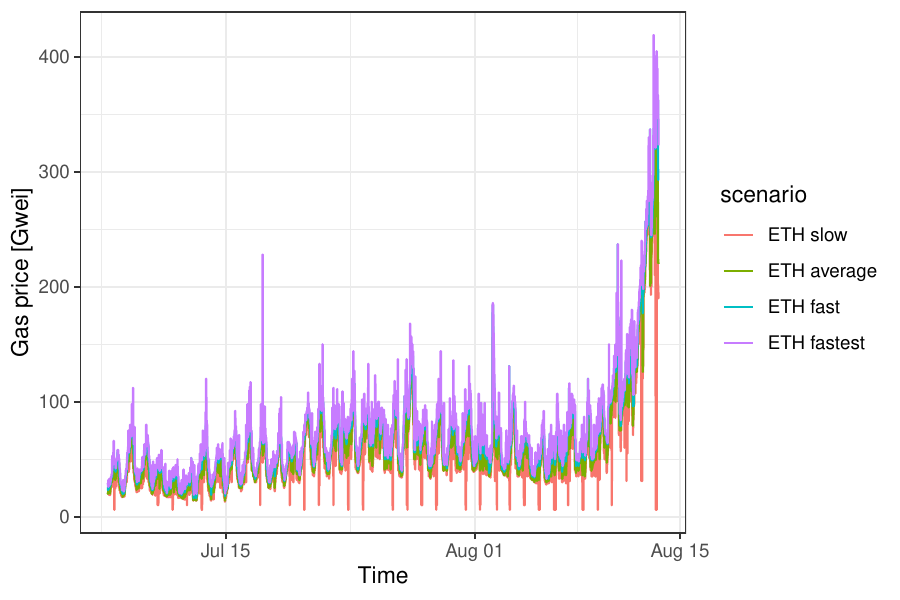}
    \caption{The price of gas required to sign the transaction on the production Ethereum blockchain under a certain time limit, based on~\cite{ETHGasSt50:online}. The data are gathered for a $\approx$37 days period between 2020-07-06 22:35 UTC and 2020-08-13 12:50 UTC. Legend: \textit{slow}~--- less than 30 minutes are required to sign the transaction, \textit{average}~--- less than 5 minutes  are needed to sign the transaction, \textit{fast}~--- less than 2 minutes are required to sign the transaction, and \textit{fastest}~--- less than 30 seconds are required to sign the transaction.}
    \label{fig:gas_price}
\end{figure}

We know that the gas price, which a user sets, controls the speed with which a transaction is processed by a miner (the higher the price is, the faster it is executed). However, how does the gas price change with time? Let us explore the empirical data\footnote{While ETH Gas Station predictions are not 100\% accurate~\cite{ap2020}; they are sufficient to illustrate the concept.} gathered from the Ethereum Gas Station service\cite{ETHGasSt50:online}, shown in Figure~\ref{fig:gas_price}.
The figure shows the price of gas a user needs to set to sign the transaction in less than 30 seconds (\textit{fastest}), 2 minutes (\textit{fast}), 5 minutes (\textit{average}), and 30 minutes (\textit{slow}). The figure also shows that the faster the transaction is, the more we have to pay. 

Our data analysis, based on the daily collection of data through the exposed API by ETH GAS Station~\cite{ETHGasSt91:online}, also shows that that the prices exhibit daily seasonality and a non-zero trend. We also see that the gas daily price is at the lowest at 23:50~UTC.

Figure~\ref{fig:cost_of_ownership} and Table~\ref{tab:cost_of_ownership} show the daily costs of the SB submissions to the Ethereum blockchain. The details of the computations are given in Appendix~\ref{sec:eth_cost_details}. As we can see, the prices will increase with the growth of the speed with which we would like the transaction to be processed, as well as the increase of intensity of the SB submissions. The daily cost of ownership ranges from \$4.07~USD for 1 transaction per day in the \textit{slow} scenario to \$2226.10~USD for 1 transaction per 5 minutes in the \textit{fastest} scenario. 

\subsubsection{IBM Blockchain}
In the case of the IBM Blockchain, we need to pay for the compute and storage resources required to process the transactions. Let us use the reference architecture recommended by the manufacturer~\cite{IBMCloud65:online}. In this case, the solution consists of the costs needed to run IBM Cloud Kubernetes cluster, IBM Blockchain Platform, and IP address allocation, as well as compute and storage resources, costing us $\approx$\$1.21~USD per hour. The cost can be reduced to $\approx$\$0.43~USD per hour by reducing computing resources. This reduction will be applicable only if the intensity of SB submission is higher than 1 SB per hour (as the resources are billable by the hour).

The cost per day will depend on the intensity of submissions of SBs and is shown in Figure~\ref{fig:cost_of_ownership} and Table~\ref{tab:cost_of_ownership}. The details of the computations are given in Appendix~\ref{sec:ibm_cost_details}. As expected, the more transactions are submitted, the higher the cost (as we need a higher amount of computing resources): the daily cost of ownership ranges from \$11.14~USD to \$29.09~USD for 1 transaction per day and per 5 minutes, respectively. 

\begin{figure}[bt]
    \centering
    \includegraphics[width=0.8\columnwidth]{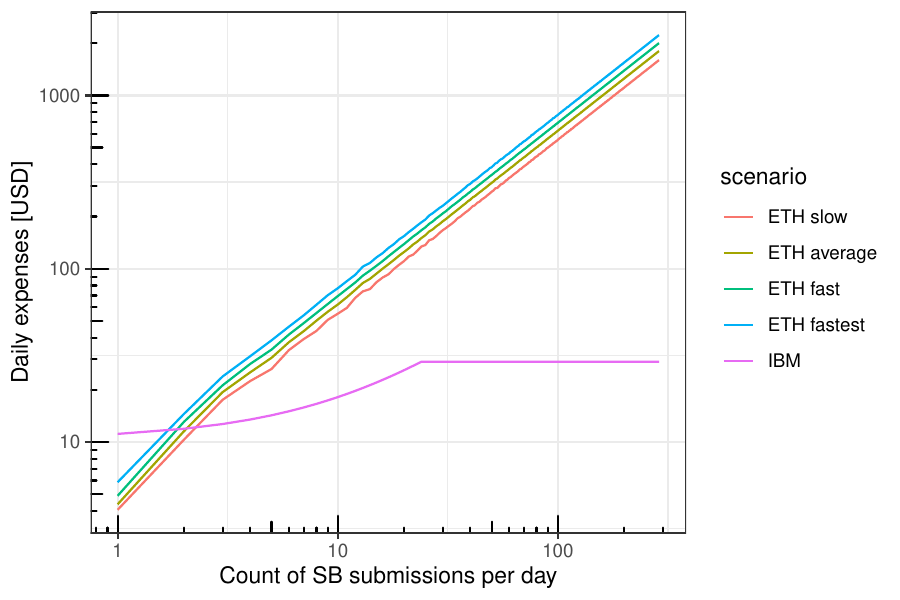}
    \caption{Daily cost of ownership. The intensity ranges from 1 SB submission per day to 288 submissions per day (i.e., one submission per five minutes).}
    \label{fig:cost_of_ownership}
\end{figure}

\begin{table}[bt]
\centering
    \caption{Daily costs for submitting SBs in USD}
    \label{tab:cost_of_ownership}

\begin{tabular}{rrrrrr}
\toprule
Submissions  & IBM & ETH & ETH & ETH & ETH \\
per day &  & slow & average & fast & fastest\\
\midrule
1 & 11.14 & 4.07 & 4.37 & 4.90 & 5.86\\
2 & 11.92 & 10.31 & 11.54 & 13.07 & 14.52\\
3 & 12.70 & 17.54 & 19.41 & 21.26 & 23.91\\
$\cdots$ & $\cdots$ & $\cdots$ & $\cdots$ & $\cdots$ & $\cdots$\\
288 & 29.09 & 1597.05 & 1800.87 & 2000.94 & 2226.10\\
\bottomrule
\end{tabular}
\end{table}

\subsubsection{IBM Blockchain vs. Ethereum blockchain}

To reiterate, the cost of ownership will depend on the intensity of the submissions of SBs. When the intensity is low, Ethereum is cheaper. When the intensity increases, IBM Blockchain becomes more economical. Quantitatively, Ethereum blockchain will be more economical if the number of SB submissions per day will be less than 2 (for the \textit{fast} and \textit{fastest} scenarios) and less than 3 (for the \textit{average} and \textit{slow} scenarios). These thresholds may change in the future (as the underlying costs are non-constant). In contrast, the underlying cost associated with running LCaaS and producing CBs is constant (the sum of computation cost for mining and the cost for storing CBs). 

Moreover, the IBM Blockchain can be used by multiple workloads. Thus, an organization may reduce the cost per workload by sharing blockchain infrastructure. 

Note that we do not consider the overhead associated with the maintenance of the IBM Blockchain infrastructure. Also, the ETH--USD exchange rate is volatile (in the last 12 months it ranged between $\approx$\$111~USD and $\approx$\$400~USD for 1~ETH~\cite{Ethereum63:online}), adding additional currency conversion risks. 

To summarize, the pros and cons are as follows. Ethereum implementation is decentralized, has higher redundancy, and a higher degree of trust (due to the large number of miners involved) than the IBM Blockchain. However, its budget is less predictable (due to fluctuation in gas price and ETH--USD conversion rate). Moreover, if the number of transactions is high, the Ethereum-based LCaaS may become prohibitively expensive.

\section{Threats to Validity}\label{sec:Threats to Validity}

In this section, we discuss the threats that affect validity structured as per~\cite{wohlin2012experimentation,yin2009case}.

\textbf{Internal validity}. Our integration of LCaaS with the blockchain vendors may be suboptimal and should be treated as a proof of idea rather than production-grade integration. Nevertheless, we were able to achieve strong performance results.

We conduct performance testing on the test network of Ethereum rather than the production network. The experiments on the test network show the computational feasibility of our approach (i.e., a transaction with SB can be submitted to a block). To extrapolate the timing results to the production network, we perform additional analysis based on the empirical data for gas prices and timing. 

In the case of the IBM Blockchain integration test, we have used the trial version of IBM Blockchain that is running on the lite infrastructure with shared resources and declared limitations~\cite{Benefits26:online}. The performance of the non-trial infrastructure may be higher due to more powerful resource allocation~\cite{Pricingf47:online}.

\textbf{Construction validity}. Manual gathering of performance data may lead to errors; thus, we automate the execution and data gathering of the experiments to minimize the risk of human errors.

\textbf{Statistical validity}. We complete several statistical tests to analyze potential factors that affect the performance of the LCaaS and its integration with blockchains. The conclusions are drawn based on the statistical validity of the tests. 

\textbf{External validity}. It is important to mention that while we have introduced additional components to LCaaS (namely AGB, RGB, TB, SB, and SBC), we have not altered the key elements of the actual blocks, so that LCaaS can be easily integrated with any other blockchain, which is proven by the successful integration of LCaaS to Ethereum and IBM Blockchain. With minor modifications, the proposed integration architecture can be used for integration between LCaaS and any other blockchain networks that supports smart contracts. Thus, LCaaS and its integration with Ethereum and IBM Blockchain can be treated as ``critical cases'' (in a case study sense~\cite{yin2009case}) showing the feasibility of our approach.

An additional threat to the validity is that one could ask why the CSCs who do not fully trust CSPs, would trust the logs provided by them? We argue that the tampering attempts typically happen after a client complains about the service. If the time difference between the submission time of logs and the rendered service is minimal, from milliseconds to a few seconds, the window of time at which the provider can tamper with the logs is narrow.

\section{Conclusion}\label{sec:Conclusion}

We described Cloud-based immutable log storage solution, LCaaS, based on the blockchain technology. This solution prevents log tampering, ensuring a transparent logging process and establishing trust between all Cloud participants (providers and users). Thus, the solution is of interest to practitioners. We provided detailed design of the solution and showed the implementation of the LCaaS on a public blockchain (Ethereum) and a private blockchain (IBM Blockchain). Performance test suggests that the solution is scalable (dealing with 100 $tps$) and is capable of fast ``sealing'' of the records (from seconds to minutes, depending on the implementation). This work is also of interest to academics, as it describes building blocks, which may be leveraged in a general scalable and immutable storage platform, leading to novel solutions for storing data.

The proposed LCaaS can act as a hierarchical ledger and a repository for all logs generated by Cloud solutions and can be accessed by all Cloud participants (namely, providers and consumers) to establish trust among them. Using verification services, a Cloud user can verify the Cloud provider's logs against the records in the hierarchical ledger and finds out if the logs were tampered with or not. 

In the future, we plan to test LCaaS with other existing blockchain solutions to find integration points that can be used to implement LCaaS on top of such solutions. Additionally, we would like to review the capacity and scalability challenges of blockchain and define key parameters that affect the LCaaS performance and cost. At the final state, we would like to use such parameters to design a framework that can reduce the total cost of ownership of an end-to-end solution for the secure storage of logs using blockchains. 

\section*{Acknowledgment}

We wish to thank the funding agencies profusely. This research is funded in part by IBM Center for Advanced Studies grant No. 1046, NSERC Discovery Grant No. RGPIN-2015-06075, and NSERC CRD Grant CRDPJ 538493--18.

\bibliographystyle{IEEEtran}
\bibliography{reference.bib}

\clearpage
\newpage

\appendix

\section{Log tampering: examples}\label{sec:examples}

Below, we provide three examples of situations where there may be motivation to tamper with logs related to private, community, and public Clouds.

\begin{example}
Case for private Cloud tamper-motivation 
In a private Cloud, a unique type of tamper-motivation may exist. Imagine a company that has established a private Cloud. The management team has requested from the Information Technology (IT) department a full second-by-second backup for all of the company's financial data. The IT department configures its backup systems and ensures that there is enough space for continuous backup of transactions. Months after the initial setup, the company's primary storage is affected by a hardware failure. The IT department finds out that the real-time backup system has stopped working a few days ago and had sent several alerts that no one in the IT department noticed. The IT department is the only department that has access to all the logs. Thus, the department may be motivated to tamper with the logs to cover up the problem.
\end{example}

\begin{example}
Case for community Cloud tamper-motivation
A community Cloud~\cite{mell2011nist} requires a clear definition of responsibilities for each partner. In case of an unfortunate incident, the party at fault may be motivated to tamper with the logs that identify them as the party responsible for the issue. Even worse, they may try to tamper with the logs and fabricate a scenario where another party becomes the main reason behind failure. Having access to the logs for one or more of the parties in a community Cloud may cause trust issues that call for the logs' immutability.
\end{example}

\begin{example}
Case for public Cloud tamper-motivation
As for the public Cloud, the platform's operational health, performance, generated metrics, and even charge-back reports (that are consolidated in the form of monthly invoices to CSCs) are entirely managed by CSPs. Without having access to the actual logs or the actual infrastructure, CSCs of the public Clouds are in a very unfair, dependent position. Consider a scenario in which a CSC deploys an application on an elastic Cloud environment with an auto-scaling feature and defines a rule that when the memory usage exceeds $80\%$, the CSP should allocate $20\%$ extra memory space to the deployed application. Imagine that the CSC receives complaints related to the application performance from its users. The CSC asks the CSP to send a detailed report of the elastic memory allocation. The CSP's IT team checks their logs and finds out that the auto-scaling feature has worked intermittently, hence the performance issue. If the CSC finds out the truth, there may be a lawsuit on the horizon. Thus, the IT team may be motivated to tamper with the log before sending it to the CSC.
\end{example}

\section{Background}\label{sec:Background}
We introduce some basic components and characteristics of blockchains in this section. In Section~\ref{subsec:Common Key Components of Blockchains}, we present a brief overview of common components of all blockchains and describe what each component does. In Section~\ref{subsec:MiningAndHashBinding}, we provide an overview of the mining and immutability of blocks in a blockchain.

\subsection{Common Key Components of Blockchains}\label{subsec:Common Key Components of Blockchains}

Here, we introduce the components that are common among all implementations of the blockchain. 

\textit{Genesis Block (GB):} Genesis block is the first block of any blockchain. Genesis block has predefined characteristics. Its \code{index} and \code{previous\_hash} are set to zero, as there are no prior blocks. The primary purpose of a genesis block is to indicate the start of a new blockchain~\cite{decker2013information}. 

\textit{Data Block (DB):} A data block, more commonly known as a block, contains the following variables: \code{index}, \code{timestamp}, \code{data}, \code{current\_hash}, and \code{previous\_hash}. The first element, \code{index}, is a unique sequential ID for each block; it uniquely identifies each block. The \code{timestamp} indicates the time at which the block is created and is usually stored in the Coordinated Universal Time format. The \code{data} is the most important element of a data block. It contains valuable information that blockchain has promised to be immutable. The \code{nonce} is an arbitrary random number that is used to generate a specific \code{current\_hash}. To achieve hash binding, each block includes a \code{previous\_hash} element. The \code{previous\_hash} is the exact duplicate of the \code{current\_hash} of the previous block. In other words, the \code{current\_hash} of the $m$-th block becomes the \code{previous\_hash} of block $m+1$. We use SHA-256 to generate \code{current\_hash} and \code{nonce} for a block as illustrated in Algorithm~\ref{alg:hash_gen}.

\begin{algorithm}[t]
\SetAlgoLined
\SetKwInOut{Input}{Input}
\SetKwInOut{Output}{Output}
\Input{\code{block\_index}, \code{timestamp}, \code{data}, \code{previous\_hash}} 
\Output{\code{current\_hash}, \code{nonce}}
 
\code{content} = concatenate(\code{index}, \code{timestamp}, \code{data}, \code{previous\_hash})\;
\code{content} = Hasher(\code{content})\tcp*{to speedup computing}
\code{nonce} = 0\;

\Repeat{prefix of current\_hash  = difficulty\_target}{
\code{nonce} = \code{nonce} + 1\;
\code{current\_hash} = Hasher( concatenate(\code{nonce}, \code{content}) )\;
}

return \code{current\_hash}, \code{nonce}\;

\caption{Generation of \code{current\_hash} and \code{nonce} for a block.}\label{alg:hash_gen}
\end{algorithm}

\textit{Blockchain (BC):}
Blocks that are linked together via hash binding will result in a blockchain. If data in an earlier block (say, block $m$) are tampered, the link among all the subsequent blocks, from $m+1$ to the most recent block $i$ will be broken. Then one has to recompute \code{current\_hash} and \code{nonce} values of each block from block $m$ to block $i$ of the BC.

\subsection{Mining and Hash Binding}\label{subsec:MiningAndHashBinding}

Relying on key characteristics of cryptographic hash function~\cite{rogaway2004cryptographic}, blocks are cryptographically sealed and mined. Mining is the process running through all possible values of an integer variable\footnote{Typically implemented as an unsigned integer.}, known as \code{nonce}, to find a value for \code{nonce}, such that if the value is added to the rest of elements of a block and then hashed, the hash matches the imposed \code{difficulty\_target}. Once the desired value for \code{nonce} is found, it resides in the \code{nonce} element of the block, and the calculated hash resides in the \code{current\_hash} element of a block. At this stage, the block is mined. The \code{difficulty\_target} is often defined as the number of required zeros at the beginning of the desired hash. The more zeros, the more computational power is needed to generate a hash that matches the difficulty target. Blocks are linked together based on a hash binding relationship. 

Once a block is mined, its content can no longer change unless the whole {PoW} process is repeated for every block in the blockchain.
In addition to hash binding, blockchains take advantage of the hash function basic properties, namely first and second preimage resistance and collision resistance, which make it extremely difficult to tamper with the hash values (ensuring their uniqueness), see~\cite{katz1996handbook} for details.

In addition to hash binding, blockchains take advantage of the hash function properties. Most cryptographic hash functions are designed to take an input of any size and produce a fixed-length hash value. Menezes et al.~\cite{katz1996handbook} indicate the following three basic properties of a hash function $h$ with inputs $x$ or $x'$ and outputs $y$ or $y'$.
\begin{enumerate}
\item 	``Preimage Resistance: for all pre\-defined outputs, it is computationally infeasible to find any input which hashes to that output, i.e., to find any preimage $x'$   
such that $h(x') = y$ for any $y$ for which a corresponding input is not known. In other words, for a given hash, it would be computationally unfeasible to reverse the hash function and find the value that was hashed.''~\cite{katz1996handbook}

\item ``Second preimage resistance: it is computationally infeasible to find any second input which has the same output as any specified input, i.e, given $x$, to find a second preimage $x' \neq x$ such that $h(x) = h(x')$.''~\cite{katz1996handbook}

\item ``Collision resistance: it is computationally infeasible to find any two distinct inputs $x$ and $x'$ which hash to the same output, i.e., such that $h(x) = h(x')$. Collision resistance implies second preimage resistance but does not guarantee preimage resistance.''~\cite{katz1996handbook}
\end{enumerate}

\subsection{Blockchain Taxonomies}\label{subsec:Blockchain Taxonomies}

Here we provide a summary of different types of blockchains and how they are compared. We will start by looking at the deployment models, including private and public blockchain and then move to the administrative models, including permissioned and permission-less blockchains. 

\subsubsection{Private and Public Blockchains}
Blockchains, as tamper-evident and tamper-resistant distributed digital ledgers, can be deployed in various ways. They can be implemented privately, inside an organization, or can be launched publicly for a wider range of customers.

In a private blockchain, nodes who can participate are often part of the same organization that controls the centralized network and offer permissions to various players. In public blockchains, since they are open to the public, all records are visible to everyone, and the network is open for new nodes to join and participate. In contrast, a public blockchain is not limited to an organization and joining the network is open to public, so is the access to data stored on blocks. 

In recent years, consortium blockchains have evolved as a hybrid model based on public blockchains. A consortium blockchain is very similar to a public blockchain except that only a group of pre-selected nodes would participate in the consensus process of the blockchain~\cite{zheng2017overview}. Table~\ref{tbl:Private-vs-Public-Blockchains} lists the differences between private and public blockchains.

\begin{table}[t]
    \centering
    \caption{Differences between Private and Public blockchains}
    \label{tbl:Private-vs-Public-Blockchains}
    \begin{tabular}{lll}
    \toprule
         & Public                  & Private               \\
    \midrule
        Network  & Decentralized           & Centralized           \\
        Security & Open Network            & Approved Participants \\
        Access   & Permission-less         & Permissioned          \\
        Identity & Anonymous or Pseudonymous & Known Identities      \\
        Speed    & Slower                  & Faster                \\
    \bottomrule
    \end{tabular}
\end{table}

\subsubsection{Permissioned and Permission-less Blockchains}
As highlighted in NIST's Blockchain Technology Overview~\cite{Blockcha24:online}, blockchain implementations are often inspired by a specific purpose or function. The purpose plays a significant role in finding and adopting the right model of blockchain. Blockchains have been categorized into two high-level categories: permission-less and permissioned (also known as permission-based). 
The permissionless blockchain allows any arbitrary user to interact with the blockchain and read and write blocks without the need of a central authority or getting approval from any party. In contrast, permission-based blockchains limit participants to specific people or organizations and provide a systematic authentication and authorization approach.

\section{LCaaS API}\label{subsec:LCaaS API}

\begin{figure*}[ht]
    \centering
    \includegraphics[width=0.8\textwidth]{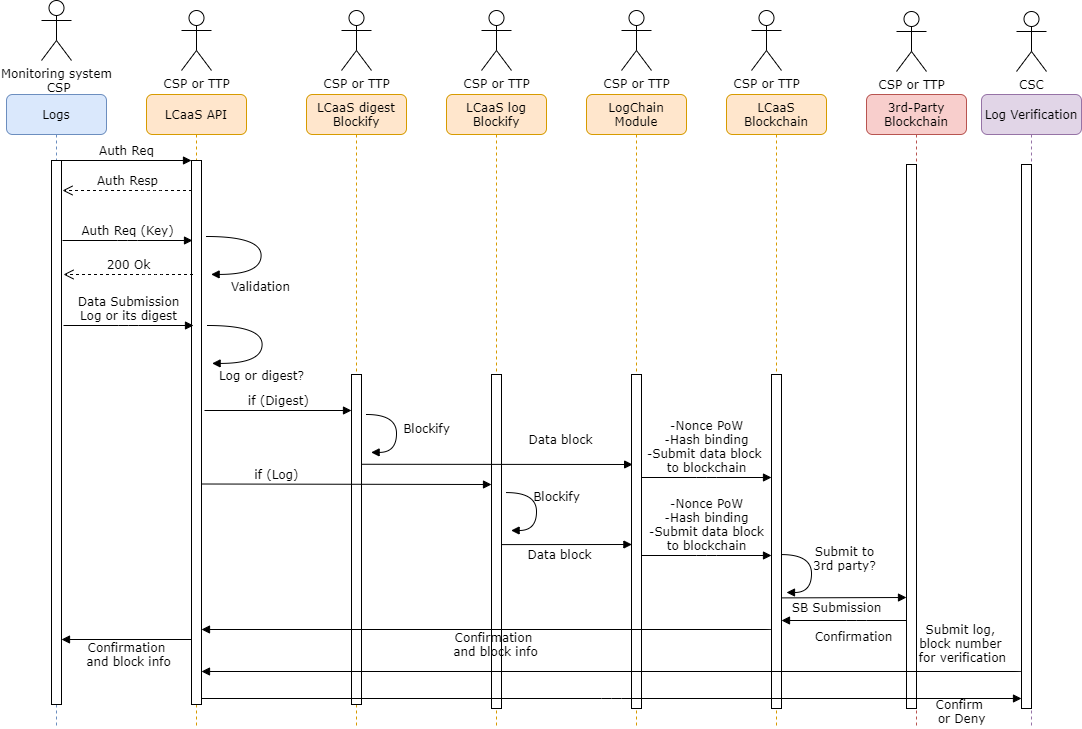}
    \caption{The Sequence Flow Diagram for LCaaS and its Internal Components}
    \label{fig:LCaaS_SequenceFlow}
\end{figure*}

As reviewed in Section~\ref{sec:Introduction}, to simplify the interaction with the LC and to allow its users to submit and verify logs easily, we introduce an API that converts the LC to the LCaaS. In other words, for each CSC, an instance of LC can be instantiated and can be used as a service through its API interfaces. Therefore, LC users do not need to integrate LC into their monitoring platform and can simply use it as a service via LCaaS API interfaces. The CSPs can efficiently use this API and interconnect the LCaaS with their monitoring systems and store all their logs, or the hash of their logs, in the Logchain. Similarly, CSCs can search and verify provided logs against the data in the Logchain and, therefore, be assured that the logs provided by the CSPs are not tampered with. However, the API is not provided to CSCs. In the current implementation, the application receives logs or their hashes, adds them to the data blocks and mines the blocks by finding a \code{nonce}. Like all other blockchains, our implementation links the blocks to their previous blocks by inserting the \code{current\_hash} of the previous block into the \code{previous\_hash} of the current block. Figure~\ref{fig:LCaaS_SequenceFlow} illustrates the sequence of actions withing the LCaaS API module from the submission of the logs to its verification. Additionally, we designate an actor for each interaction.

\subsection{Submission Methods}

The API is designed and implemented using Flask~\cite{flask}, a micro-framework for Python. There are two data submission methods: \code{submit\_raw} and \code{submit\_digest}. The former allows the client to submit the actual logs, while the latter~--- just the file's digest (e.g., SHA-based digest computed using OpenSSL \code{dgst}~\cite{openssl}), thus, preserving the privacy of the log and reducing the amount of transmitted data. Both methods return, on success, \code{timestamp}, \code{block\_index} and other details of the created block and, on failure, details of the error. For privacy reasons, the CSPs or CSCs may decide to generate a digest locally and submit it to LCaaS by using the \code{submit\_digest} method. The Client of the LCaaS, have full control over the amount of log included in each submission and their interval, Therefore, LCaaS API can be really customized based on the presented scenario and the CSPs and CSCs needs.

\subsection{Verification Methods} %\label{subsec:verfication}

There are three verification methods: \code{verify\_raw}, \code{verify\_digest}, and \code{verify\_tb}. The first one allows the client to verify the existence of actual logs in the LCaaS. The second one allows the client to verify the digest of the logs. The last method allows a user to verify the existence of a terminal block with a specified hash in the LCaaS, hence, proving the integrity of all the blocks in the circled blockchain of the submitted terminal block with one verification operation. All methods return, on success, the details of the found values in the LCaaS and, on failure, details of the error.

For the verification of the actual log content, one should use method \code{verify\_raw}. The method would return the status of submission and number of blocks that match the submitted data; if no block is found, the API will return a message informing the user that no match has been found. In case of an error, the API will return the failed status along with the error's description.

To improve the scalability of our solution, we introduce the \code{verify\_tb} method. It provides an assurance (in the cryptographic sense~\cite{rogaway2004cryptographic}) that the sequence of blocks, from \code{index\_from} to \code{index\_to} are not tampered with. By comparing the generated hash value from all the \code{current\_hash} values of a circled blockchain to the \code{aggr\_hash} value in the data element of a TB, one can verify the integrity of all the blocks in the circled blockchain.

It is important to mention that while we have introduced these additional components, we have tried not to alter the key element of the actual blocks. Avoiding any alteration on block's structure is intentional, because any modification in the blocks format (e.g., adding new elements) will result in a proprietary implementation of blocks and blockchains and will reduce the applicability of the proposed hierarchical structure to other existing blockchain platforms.

\section{Hierarchical Structure of LCaaS}\label{subsec:Hierarchical Structure of LCaaS}

\begin{figure*}[ht]
\begin{center}
\includegraphics[width=0.75\textwidth]{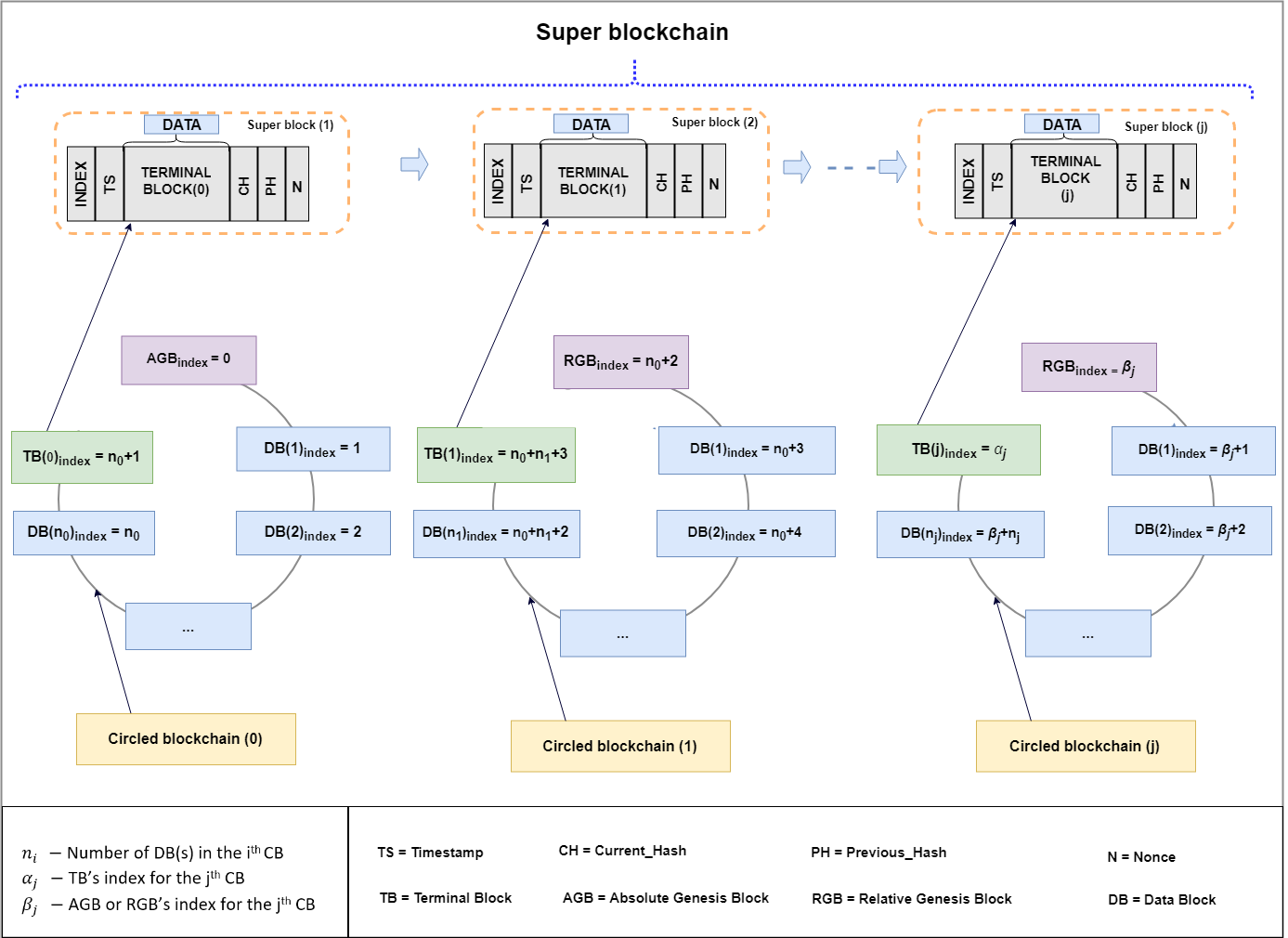}
\caption{Two-level Hierarchy as implemented by LCaaS}
\label{fig:two_level_hierarchy}
\end{center}
\end{figure*}

To overcome blockchain performance issues, LCaaS uses a two-level hierarchy, but the number of levels can be increased if a use case requires it, making it a cascading blockchain. Figure~\ref{fig:two_level_hierarchy} depicts this hierarchy and the internal relationship among its components.

As mentioned in the legend of Figure~\ref{fig:two_level_hierarchy}, $n_i$ refers to the number of data blocks in the $i$-th circled blockchain and $\alpha_j$ is the index of the terminal block of the $j$-th circled blockchain. $\beta_j$ is the index for the absolute or relative genesis block of the $j$-th circled blockchain. The value of $\alpha_j$ will be calculated by

\begin{equation}
    \alpha_j = 
\begin{cases}
    n_0 + 1,& \text{if } j = 0\\
    n_0 + 1 +  \sum_{i=1}^{j} {(n_i + 2)}  {n_i},             & \text{if } j\geq 1
\end{cases}
\end{equation}
and the value of $\beta_j$ will be calculated by
\begin{equation}
    \beta_j = 
\begin{cases}
    0,& \text{if } j = 0\\
    \alpha_{j-1} + 1,              & \text{if } j\geq 1
\end{cases}.
\end{equation}

\begin{figure}[htb]
\begin{center}
\includegraphics[width=\columnwidth]{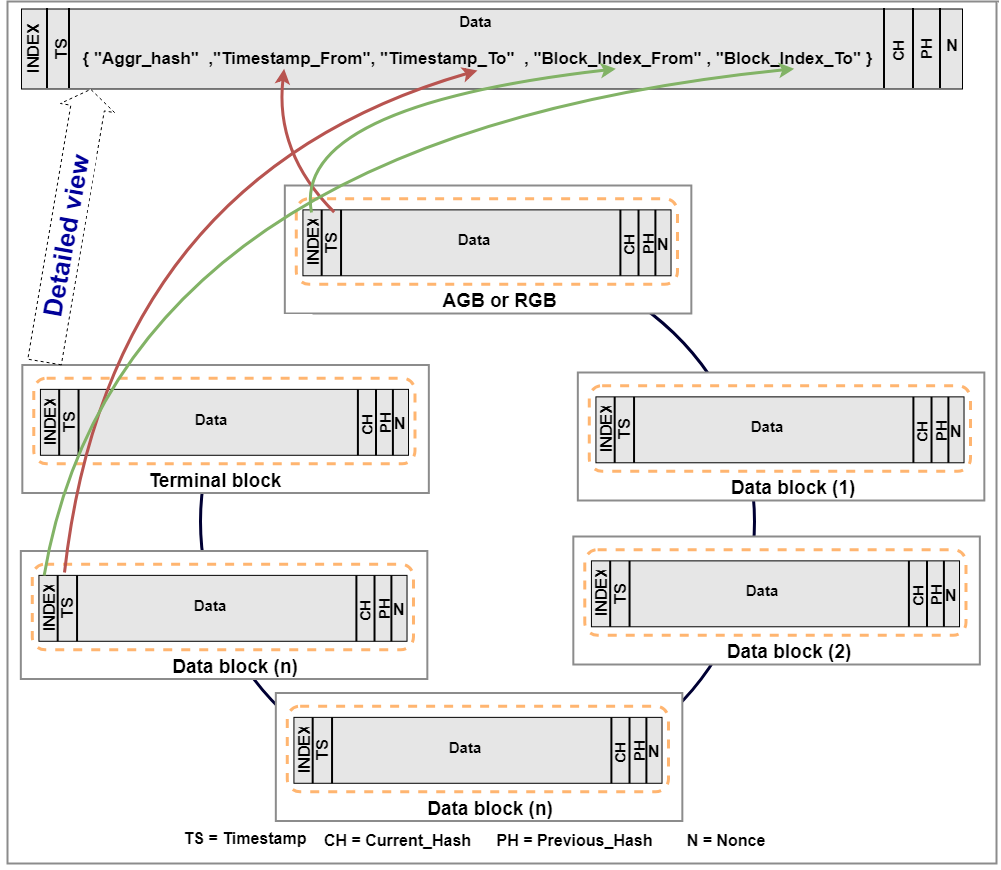}
\caption{The Relation Between Terminal Block and All other Blocks in a Circled Blockchain}
\label{fig::TerminalBlockdetails}
\end{center}
\end{figure}

Figure~\ref{fig::TerminalBlockdetails} shows the relationship between a terminal block and all other blocks in a circled blockchain.

\begin{figure}[htb]
\begin{center}
\includegraphics[width=\columnwidth]{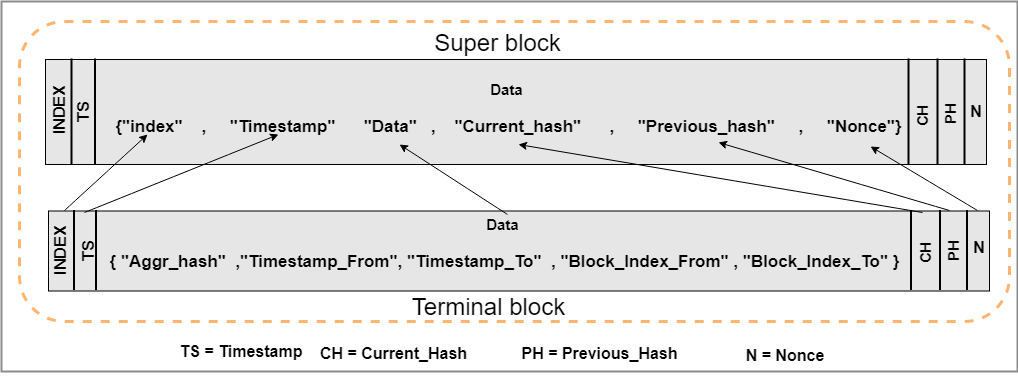}
\caption{The Relationship Between Terminal Block and Super Block}
\label{fig::SBTB}
\end{center}
\end{figure}

Figure~\ref{fig::SBTB} depicts the relationship between a TB and the data element of a SB. 

\section{Details of computing the cost of SB submission}\label{sec:cost_details}
Details of computing the cost of SB for IBM Blockchain are given in Appendix~\ref{sec:ibm_cost_details}, for Ethereum~--- in Appendix~\ref{sec:eth_cost_details}.

\subsection{IBM Blockchain}\label{sec:ibm_cost_details}
A typical small-scale, production-grade setup of the IBM Blockchain costs \$1.19 USD per hour~\cite{IBMCloud65:online}. The recommended setup requires three VMs (two for the peers to ensure high availability, and one to act as a Certificate Authority), an access to a production IBM Blockchain Platform, storage, and a Kubernetes cluster (to tie it all together).

To the variable costs above, we also need to add IP address allocation cost of \$16.00 USD per month (or $\approx 16.00 / 30 / 24 = $ \$0.02 USD per hour. This will increase hourly cost to $\approx$ \$1.21~USD per hour.

When the system is inactive, we can scale down the CPU allocation to almost zero to reduce the cost~\cite{IBMCloud65:online}. In this case, the hourly price will be reduced to \$0.43~USD per hour.

Assuming that we would like to equidistribute SB submissions around the clock-dial, we can take the advantage of the cost reduction. The cost in this case will be computed as per Algorithm~\ref{alg:ibm_daily_cost}.

\begin{algorithm}[ht]
\SetAlgoLined
\SetKwInOut{Input}{Input}
\SetKwInOut{Output}{Output}
\Input{$s$ \tcp*{SB count per day}} 
\Output{$d$ \tcp*{Daily cost in USD}}

$c$ = $16.00 / 30 / 24$ \tcp*{IP address hourly cost}
$f$ = $1.19 + c$ \tcp*{Full cost per hour}
$r \approx 0.41 + c$ \tcp*{Reduced cost per hour}

\eIf{$s < 24$}{
   $d$ = $s \cdot f + (24 - s) \cdot r$\;
   }{
   $d$ = $24 \cdot f$\;
  }

return $d$\;

\caption{Computation of daily cost for IBM Blockchain.}\label{alg:ibm_daily_cost}
\end{algorithm}

\subsection{Ethereum Blockchain}\label{sec:eth_cost_details}
We have gathered Ethereum gas prices every 5 minutes for 37 days. The results are shown in Figure~\ref{fig:gas_price}. 

\begin{figure*}[ht]
    \centering
    \includegraphics[width=\textwidth]{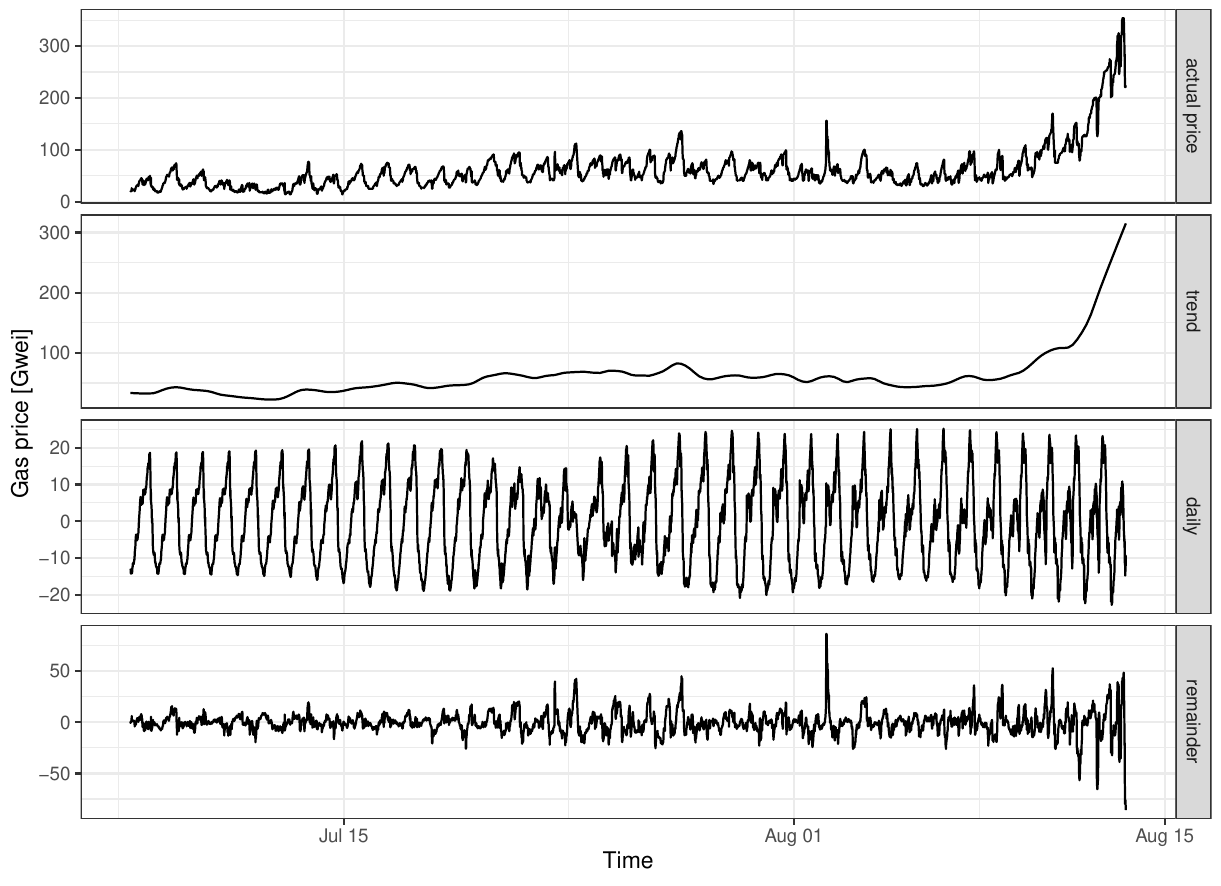}
    \caption{Example of gas price decomposition (\textit{average} scenario) performed using R forecast package~\cite{Hyndman08,Hyndman20}.}
    \label{fig:avg_gas_price_decomposition}
\end{figure*}

These empirical data are passed to Algorithm~\ref{alg:eth_daily_cost} to compute daily price for a given scenario and SB submission intensity. In a nutshell, the algorithm works as follows. Based on our analysis (example of gas prices time series decomposition is shown in Figure~\ref{fig:avg_gas_price_decomposition}), the lowest ETH gas price is, consistently, at 23:50~UTC. Thus, this will be our starting time. Similar to the approach in Appendix~\ref{sec:ibm_cost_details}, we will equidistribute SB submissions around the clock-dial. For example, if we need to submit one SB per day the submission will happen at 23:50~UTC. In the case of two SB submissions, they will be sent off at 23:50~UTC and 11:50~UTC, in the case of three SB submissions --- at 23:50~UTC, 7:50~UTC, and 15:50~UTC, and so on. For each scenario, we will then average out gas prices for a given minute of the day and sum them up. To convert this value into US dollars, we will need to multiply the average gas price by the number of units of gas needed to execute SB submission contract (namely, $\approx$335K gwei) and then convert gwei to USD. In our case, we will use the average ETH-USD exchange rate between 2020-07-06 and 2020-08-13 (i.e., the time interval when the empirical gas prices were collected), which was equal to 302.05 USD per 1 ETH~\cite{Ethereum63:online} or $3.02 \times 10^{-7}$~USD per 1 gwei.

\begin{algorithm}[ht]
\SetAlgoLined
\SetKwInOut{Input}{Input}
\SetKwInOut{Output}{Output}
\Input{$s, r, p, c$ \tcp*{SB count per day, the name of the Ethereium scenario, spot prices of gas in gwei, and gwei to USD conversion rate, respectively}} 
\Output{$d$ \tcp*{Daily cost in USD}}

Transform $p$ to retain three columns for each observation: \code{minute\_of\_the\_day}, \code{scenario\_name}, \code{gas\_price}.\;

Retain the observations in $p$ only for the scenario $r$\;

Perform the following aggregation: ``select \code{minute\_of\_the\_day}, average(\code{gas\_price}) from $p$ group by {minute\_of\_the\_day}''\;

For a given $s$, equidistribute the submission times, starting at 23:50 UTC and identify the closest \code{minute\_of\_the\_day} to each submission time\;

$z$ = Sum up the average \code{gas\_price} for each of the closest \code{minute\_of\_the\_day}\;

$g$ = $335000$  \tcp*{Number of gas units needed for SB submission (measured in gwei)}

$d$ = $z \cdot g \cdot c$\;

return $d$\;

\caption{Computation of daily cost for Ethereum Blockchain.}\label{alg:eth_daily_cost}
\end{algorithm}

\end{document}